\documentclass[aip,jap]{revtex4-1}
\usepackage{color,epsfig}
\usepackage{bm}
\usepackage{amsmath,amsfonts,amssymb,bm}
\usepackage{hyperref}
\usepackage{subfigure}

\newcommand{\bea}{\begin{eqnarray}}
\newcommand{\eea}{\end{eqnarray}}
\newcommand{\be}{\begin{equation}}
\newcommand{\ee}{\end{equation}}

\newcommand\bmb{\left( \begin{matrix}}
\newcommand\emb{\end{matrix} \right)}
\renewcommand{\(}{\left(}
\renewcommand{\)}{\right)}

\renewcommand\vec{\bm}

\begin{document}
\title{Thermoelectric transport in torsional strained Weyl semimetals}

\author{Enrique Mu\~{n}oz}
\affiliation{Facultad de F\'isica, Pontificia Universidad Cat\'olica de Chile, Vicu\~{n}a Mackenna 4860, Santiago, Chile}

\author{Rodrigo Soto-Garrido}
\affiliation{Facultad de Ingenier\'ia y Tecnolog\'ia, Universidad San Sebasti\'an, Bellavista 7, Santiago 8420524, Chile}

\date{\today}

\begin{abstract} In a recent paper \cite{Soto_018} we studied the electronic transport properties in Weyl semimetals submitted to the combined effects of torsional mechanical strain and magnetic field, showing that this configuration induces a node-polarization effect on the current that can be used to measure the torsion angle from transmission experiments. In this article, we extend our previous work to study thermoelectric transport in Weyl semimetals (WSM) under torsional strain and an external magnetic field. 
Our analysis involves only the electronic contribution to the transport coefficients, and is thus valid at low temperatures where the phonon contribution is negligible. 
We provide exact analytical expressions for the scattering cross section and the transmitted heat current, in order to calculate the thermal conductance and  the Seebeck coefficient under this configuration. Our results suggest
that thermoelectric transport coefficients in these materials can be engineered by appropriately tuning the magnitude of the magnetic field, torsional strain and the applied bias or thermal gradient, leading to a potentially very high figure of merit.
\end{abstract}

\maketitle

\section{Introduction}
\label{sec:Intro}

In recent years,  the discovery of a class of materials that exhibit strong spin-orbit coupling and hence nontrivial topological properties  \cite{armitage2017,Hasan_017,Yan_017,Burkov_016} (for instance, the magnetic pyrochlore iridates materials) lead to a solid-state physics scenario for a former theory developed by Weyl in the 1920's \cite{Weyl_29}. The so-called Weyl fermions, in addition to spin, possess an intrinsic property called chirality (the projection of spin over the momentum direction). Remarkably, this quantum mechanical property induces novel macroscopic effects in the transport phenomena on these materials.

In topological Weyl semi-metals (WSM), the strong spin-orbit coupling induces time reversal symmetry or inversion symmetry, thus leading to the splitting of a Dirac point in the spectrum into a pair of Weyl nodes of opposite chirality. 

Despite their Dirac-like dispersion relation in the vicinity of the metallic points, WSM do not necessarily satisfy Lorentz covariance. For instance, type II WSM
violate this condition \cite{Hasan_017,armitage2017,Burkov_016,Yan_017}, due to a dispersion relation where the Dirac cone is strongly tilted with the consequent formation of electron and hole pockets \cite{Huang_016}. Among the important consequences of this effect, that are verifiable experimentally, is the chiral anomaly \cite{Morimoto_016} associated to a strong dependence of the electronic transport response on the direction of the applied electric field. The effects of mechanical strain on the transport properties of Dirac-type materials, while extensively studied in graphene (see for instance Refs. \cite{Amorim_016,Naumis-2017} and references therein), are still a matter of debate for WSMs. As in the case of graphene, different types of elastic strains give rise to different types of induced gauge fields in WSM \cite{Cortijo_015,Cortijo-2016,vozmediano2017-1}. These induced gauge fields can manifest themselves in the chiral anomaly effect\cite{Pikulin-2016}, in the energy spectrum \cite{Grushin-2016}, in quantum oscillations in the absence of a magnetic field \cite{Liu-2017}, in the collapse of the Landau levels \cite{Arjona-2017} and, as recently
shown by us, in a node polarization effect over the electric current \cite{Soto_018}. Thermoelectric transport in WSM is particularly interesting, in part since
the linear Dirac-type dispersion relation induces a non-trivial dependence on the chemical potential \cite{Lundgren_014}. Moreover, thermoelectric signatures of the chiral
anomaly have been experimentally observed\cite{Jia_016} in $Cd_3 As_2$. On the more practical side, there are strong experimental indications that WSM under external magnetic fields
may constitute attractive materials for thermoelectric applications, with a potentially very high figure of merit\cite{Skinner_018} $ZT \sim 10$.

In this work, we focus on the combined effects of torsional mechanical strain and an external magnetic field on the thermoelectric transport properties in WSM, using an analytic procedure developed in Refs. \cite{Munoz-2017,Soto_018} by ourselves. In a previous study \cite{Soto_018}, we
demonstrated the node-filtering effect of the electronic current that can be achieved through a combination of torsional strain and an external magnetic field in a type I WSM. In this work, we extend our previous results to show that it is possible to engineer, by means of a combination of torsional strain, magnetic field and bias, the
thermoelectric coefficients in these materials, such as the thermal conductance and the Seebeck coefficient. Our analysis involves only the electronic contribution to the transport coefficients, and is thus valid at very low temperatures where the phonon contribution is negligible.

\section{Model} 
\label{sec:model}
One of the simpler cases of a WSM, is a type I WSM with two nodes of opposite chirality. In this case, the low energy Hamiltonian is given by a block Hamiltonian around the two nodes located at  $\mathbf{K}_{\pm}=\pm\vec{b}/2$, where $\vec{b}$ is the vector between the two nodes in momentum space and $v_F$ is the Fermi velocity \cite{vozmediano2017-1}:
\begin{equation}
\hat{H}_{b}=\bmb \hat{H}_{0}^{+} & 0\\
 0 & \hat{H}_{0}^{-} 
 \emb=v_F\bmb \vec{\sigma}\cdot\hat{\vec{p}} & 0\\
 0 & -\vec{\sigma}\cdot\hat{\vec{p}} 
 \emb
\end{equation}
where $\vec{\sigma}=(\sigma_1,\sigma_2,\sigma_3)$ are the Pauli matrices and $\hat{\mathbf{p}}$ is the momentum operator.
The values of $b$ and $v_F$ depend on the specific material. 

\subsection{Strain}
\label{sec:strain}

In a quite recent paper Arjona and Vozmediano \cite{vozmediano2017-1} showed how rotational strain induces pseudo-gauge fields in the low energy effective theory. Closely following their work [\onlinecite{vozmediano2017-1}], we write the effective model for the case of torsional strain.  Assuming that the two Weyl nodes are separated by a vector in the $z$-direction $\vec{b}=b\hat{z}$ and that the cylinder has a length $L$ in the $z$-direction, the displacement vector $\mathbf{u}$  \cite{vozmediano2017-1} gives rise to $\vec{\Omega}=\frac{1}{2}\(\vec{\nabla}\times\vec{u}\)$, in terms of which the pseudo-vector potential associated to the deformation is given by:
\begin{equation}
 \vec{A}=\vec{b}\times\vec{\Omega}=\frac{\theta b}{2L}\(-y\hat{x} +x\hat{y}\)
\end{equation}
and the associated induced pseudo-magnetic field is therefore:
\begin{equation}
 \vec{B}_s=\(\vec{\nabla}\times\vec{A}\)=\frac{\theta b}{L}\hat{z}.
 \label{eq:Bs}
\end{equation}
where $\theta$ is the torsion (twist) angle.
An important point to remark is that the pseudo-magnetic field possesses opposite signs at each of the two nodes\cite{Cortijo_015,Cortijo-2016b}. In contrast, a real magnetic field preserves the same sign at both nodes. Closely following our previous analysis in Ref.\cite{Soto_018}, besides the  torsional strain we shall also consider the presence of an external magnetic field that gives rise to the node-polarization effect on the electric current \cite{Soto_018} (in analogy to the valley-polarization effect observed in graphene \cite{Munoz-2017}).

\section{Scattering through a cylindrical region with magnetic field and mechanical strain}
\label{sec:scattering}

\begin{figure}[hbt]
\centering
\includegraphics[width=0.6\columnwidth]{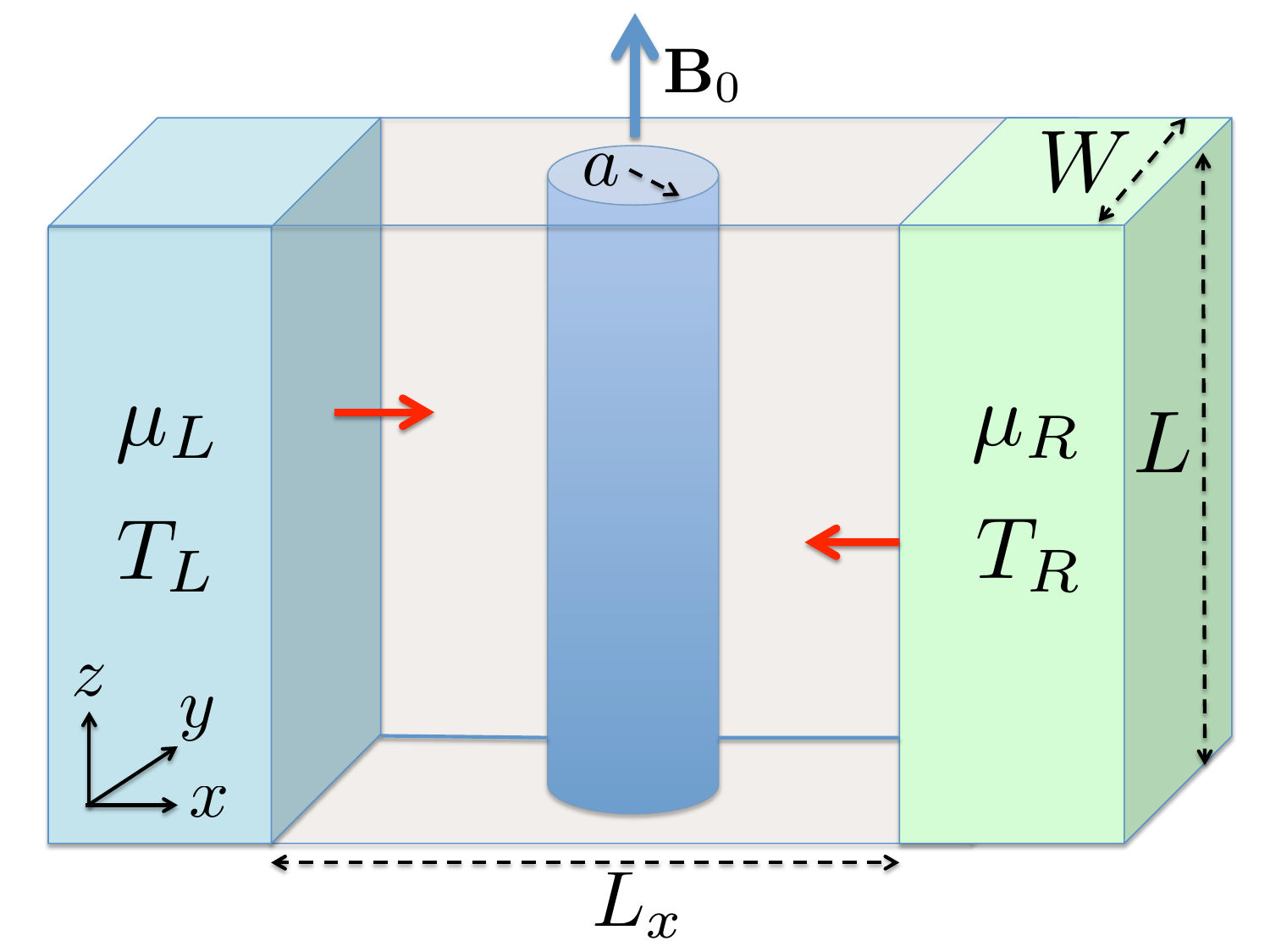}
\caption{(Color online) Pictorial description of the system under consideration: A WSM slab of dimensions $L\times W \times L_x$, with a cylindrical region of radius $a$ submitted to a combination of torsional strain and an external magnetic field $\vec{B}=B_0\hat{z}$.}
\label{fig1}
\end{figure}

For the sake of completeness, in this section we review the scattering analysis proposed for us in references \cite{Munoz-2017,Soto_018}. We will consider the problem of three-dimensional elastic scattering of an incident free spinor with momentum $\mathbf{k} = (k_x,k_y,k_z)$ and energy $E_{k,\lambda}^{\xi} = \lambda \hbar v_F |\mathbf{k}|$, where $\lambda = \pm 1$ is the ``band'' index and $\xi = \pm 1$ labels each of the Weyl nodes located at $\mathbf{K}_{\xi}=\xi\vec{b}/2$. Due to the translation symmetry in the $z$-direction, $k_z$ is a good quantum number and it is possible to decouple the $z-$direction from the plane and therefore the analysis in \cite{Munoz-2017,Soto_018} is applied. Moreover,
we shall specialize our scattering analysis to the geometry shown in Fig.\ref{fig1}, depicting a nano-junction where the WSM cylindrical region is located in between
two contacts whose surface normal points in the x-direction. Therefore, we are interested in transport due to counter-propagating waves in the x-direction and hence
we set $k_y = 0$.

We start by considering a free fermion propagating towards the cylindrical scattering center, which is described by the eigenvector of the equation
\begin{eqnarray}
\hat{H}_{0}^{\xi} \tilde{\Psi}_{in,\mathbf{k}}^{(\lambda,\xi)}(r,\phi,z) = \lambda \hbar v_F |\mathbf{k}| \tilde{\Psi}_{in,\mathbf{k}}^{(\lambda,\xi)}(r,\phi,z), 
\end{eqnarray}
where free spinor is given by:    
\begin{equation}
\tilde{\Psi}_{in,\mathbf{k}}^{(\lambda,\xi)}(r,\phi,z) = \frac{1}{\sqrt{1 + \left(\frac{\lambda\xi|\mathbf{k}| - k_z}{k_x} \right)^2}}\left(\begin{array}{c}1 \\ \frac{\lambda\xi|\mathbf{k}| - k_z}{k_x} \end{array} \right) e^{i k_x r\cos\phi+ik_z z}.
\label{eq:incm}
\end{equation}
We now apply the standard partial wave analysis for scattering used in \cite{Munoz-2017,Soto_018} and described in most textbooks (see for instance, Ref. \cite{sakurai}). Using the symmetry around the $z-$axis, we can choose as a quantum number the eigenvalue $\hbar m_j$ of the total angular momentum operator in the $z-$direction ($\hat{J}_3 = \hat{L}_3 + \hat{\sigma}_3/2$) to label the solutions:
\begin{equation}
\tilde{\Psi}_{m_j,k_z}^{(\lambda,\xi)}(r,\phi,z) = r^{-1/2}\left(\begin{array}{c} f_{m_j}(r) e^{i(m_j -1/2)\phi}\\-i\,g_{m_j}(r) e^{i(m_j + 1/2)\phi}
\end{array}\right)e^{ik_zz}. 
\label{eq:free}
\end{equation}

As explained in Refs. \cite{Soto_018,Munoz-2017} we find:
\begin{align}
f_{m_j}(r) =& c_1 \sqrt{k_x r} J_{m_j -1/2}(k_x r) + c_2 \sqrt{k_x r} Y_{m_j - 1/2}(k_x r),\nonumber\\
g_{m_j}(r) =& c_3 \sqrt{k_x r} J_{m_j+1/2}(k_x r) + c_4 \sqrt{ k_x r} Y_{m_j+1/2}(k_x r),
\label{eq:bessel}
\end{align}
where the coefficients $\{c_j\}$ are determined from the appropriate boundary and asymptotic conditions as was shown in detail in Refs.\cite{Munoz-2017,Soto_018}. In addition, as in standard elastic scattering theory \cite{sakurai}, the phase shift captures the effect of a scattering region over the transmitted particle waves.  

To calculate the phase shift $\delta_m$ associated to each angular momentum channel $m \equiv m_j - 1/2$, we need to match each spinor component of the free solution Eq. \eqref{eq:free}, and its first derivative, to the solution inside the region submitted to the effective magnetic field $B_\xi=B_0+\xi B_S$ (where $B_0$ is the external magnetic field and $B_s=\theta b/L$ is the pseudo-magnetic field induced by the torsional strain, Eq. \eqref{eq:Bs}) at the boundary of the cylinder $r=a$. The spectrum inside the cylindrical region corresponds to relativistic pseudo Landau levels, as seen in Refs.~\cite{Munoz-2017,Soto_018}, with an effective pseudo magnetic field that is node-dependent,
\begin{eqnarray}
E_{\lambda}^{\xi}(n) = \lambda \hbar v_F \sqrt{2 n |B_{\xi}|/\tilde{\phi}_0  + k_z^2}, 
\label{eq_spectrum}
\end{eqnarray}
with $\tilde{\phi}_0 = \hbar v_F/e$ the flux quantum defined in terms of the Fermi velocity in the WSM.
As it was shown in Refs. \cite{Munoz-2017,Soto_018},the expression for the phase shift $\delta_m$ can be obtained analytically and is given by:
\begin{widetext}
\begin{equation}
\tan\delta_m =  \frac{J_{m+1}(k_x a) + \displaystyle\frac{J_{m}(k_xa)}{k_x a}\left\{|m| - m - \frac{|B_{\xi}|a^2}{2\tilde{\phi}_0}- \frac{L_{n_{\rho}-1}^{|m|+1}(|B_{\xi}|a^2/2\tilde{\phi}_0)}{L_{n_{\rho}}^{|m|}(|B_{\xi}|a^2/2\tilde{\phi}_0)}
\right\}}
{Y_{m+1}(k_x a) + \displaystyle\frac{Y_{m}(k_x a)}{k_x a}\left\{
|m| - m - \frac{|B_{\xi}|a^2}{2\tilde{\phi}_0}- \frac{L_{n_{\rho}-1}^{|m|+1}(|B_{\xi}|a^2/2\tilde{\phi}_0)}{L_{n_{\rho}}^{|m|}(|B_{\xi}|a^2/2\tilde{\phi}_0)}
\right\}}.
\label{eq:phase}
\end{equation}
\end{widetext}
where we have defined $m \equiv m_j - 1/2$.
 
\subsection{Scattering cross section}
Well outside the cylindrical region where the effective magnetic field is present ($r\gg a$), the state is a linear combination of the incident and scattered spinors \cite{Munoz-2017,Soto_018}:
\begin{equation}
\tilde{\Psi}_{out}(r,\phi,z) \sim\frac{1}{\sqrt{1 + \left(\frac{\lambda\xi|\mathbf{k}| - k_z}{k_x} \right)^2}}\left(\begin{array}{c}1 \\ \frac{\lambda\xi|\mathbf{k}| - k_z}{k_x} \end{array} \right) e^{i k_x r\cos\phi+ik_ z} + \left( \begin{array}{c}f_{1}(\phi)\\f_{2}(\phi)\end{array} \right)\frac{e^{ik_x r + i k_z  z}}{\sqrt{r}},
\label{eq:out}
\end{equation}
with amplitudes $f_1(\phi)$ and $f_2(\phi)$ for each component of the scattered spinor. Moreover, this expression must be equal to the asymptotic form of the solution for $r\gg a$, represented in terms of
phase shifts (see the specific details in Refs. \cite{Munoz-2017,Soto_018}). 

When $L \gg 1/k_F$ (a long cylinder), the differential scattering cross-section per unit length is given by \cite{Soto_018},
\begin{equation}
\frac{d\tilde{\sigma}}{d\phi} = |f_1(\phi)|^2 + |f_2(\phi)|^2 = \frac{2}{\pi k_x}\frac{|\vec{k}|+\lambda\xi k_z}{|\vec{k}| - \lambda\xi k_z}\sum_{m,m'} e^{i(\delta_m - \delta_{m'})}\sin\delta_m \sin\delta_{m'} e^{i(m - m')\phi}
\label{eq:cross_section}
\end{equation}
and the total scattering cross section is then given by integrating over the scattering angle $\phi$ ($0\le \phi \le 2\pi$) and over the length of
the cylinder \cite{Soto_018}
\begin{align}
\sigma =& \int_{0}^{L} dz \int_0^{2\pi} d\phi\, \frac{d\tilde{\sigma}}{d\phi} = \frac{4 L}{k_x}\frac{|\vec{k}|+\lambda\xi k_z}{|\vec{k}| - \lambda\xi k_z}\sum_{m=-\infty}^{\infty}\sin^2\delta_{m}.
\label{eq_total_scatt}
\end{align}

An important point is that the assumption of a very long cylinder, $L \gg 1/k_F$, is well founded. In the case of TaAs, where $b\sim 0.08$ $\mathring{\text{A}}^{-1}$ and $v_F\sim 1.3\times 10^{5}$ m/s, $1/k_F\sim 9$ $\mathring{\text{A}}$, so even a slab of a few microns is already in the range of validity of our expressions. 
In addition, for Cd$_3$As$_2$, $b\sim 0.2$ $\mathring{\text{A}}^{-1}$ and $v_F\sim 1.5\times 10^{6}$ m/s, $1/k_F\sim 0.8$ $\mathring{\text{A}}$ \cite{neupane-2014}
and hence the applicability of our expressions is even more striking in this second example.

\section{Transmission and Landauer ballistic current}

In the following, we will study the transport properties of a slab of a WSM of height $L \gg 1/k_F$  ($z$-direction), width $W \gg 1/k_F$ ($y$-direction) and
length $L_x \gg 1/k_F$ ($x$-direction), which is connected to two semi-infinite WSM contacts held at different chemical potentials $\mu_L$ and $\mu_R$, respectively. In addition, a uniform magnetic field in the $z-$direction ($B_0\hat{z}$) and torsional mechanical strain is applied inside a cylindrical region of radius $a$. The application of torsional strain will induce a pseudo-magnetic field in the $z-$direction as it was showed in sec. \ref{sec:strain}. Using the Landauer ballistic approach\cite{Datta_1,Datta_2,Nazarov}, the net current along the slab ($x$-direction) is given by the net counterflow of the particle currents emitted from the left and right semi-infinite WSM contacts, respectively. Each contact is assumed to be in thermal equilibrium, with the Fermi-Dirac distributions $f(E-\mu_L,T_L)\equiv f_L(E)$ and  $f_R(E) \equiv f(E-\mu_R,T_R)$, respectively.  A schematic of the system is shown in Fig.~\ref{fig1}.


The effect of the external magnetic field and the pseudo-magnetic field over charge transport can be expressed as an effective  cross-section $W L T_{\xi}(E,\phi)$, with $T_{\xi}(E,\phi)$ the transmission coefficient\cite{Datta_1,Datta_2,Nazarov} in the direction specified by the polar angle $\phi$, for an incident spinor arising from the node $\mathbf{K}_{\xi}$. We thus define the effective cross-section in the $\phi$-direction by \cite{Munoz-2017,Soto_018}:
\begin{align}
W L\, T_{\xi}(E_k,\phi) &= \sum_{n,\lambda}\frac{L}{\tilde{\sigma}(\mathbf{k})} \frac{d\tilde{\sigma}(\mathbf{k})}{d\phi}  \delta\left(\lambda |\vec{k}| - \frac{E_{\lambda}^{\xi}(n)}{\hbar v_F}\right),
\label{eq_transm}
\end{align}
with $E_{\lambda}^{\xi}(n)$ the energy spectrum inside the cylindrical region, as defined in Eq.(\ref{eq_spectrum}). For the remaining of this work, we will considerer an incident spinor with a wave vector $\vec{k}=(k_x,0,0)$, i.e. a plane wave propagating in the $x-$direction. 

The net electric current flowing across the region will be $I = I_{+} + I_{-}$, with the node component given by \cite{Munoz-2017,Soto_018}
\begin{eqnarray}
I_{\xi} = e v_F W L  \int_{-\infty}^{\infty} dE \left[ D_L(E)f_L(E) - D_R(E)f_R(E)\right] \bar{T}_{\xi}(E),
\label{eq_current0}
\end{eqnarray}
where $D_{L/R}(E)$ is the density of states at each contact. Here,  we have defined the net transmission coefficient for Dirac spinors at node $\mathbf{K}_{\xi}$ as the angular average  $\bar{T}_{\xi}(E) = \int_{-\pi/2}^{\pi/2}d\phi\,\cos\phi\,T_{\xi}(E,\phi)$.
Assuming that both contacts are identical semi-infinite regions, the density of states are equal, and given by
\begin{eqnarray}
D_L(E) &=& D_R(E) = \frac{2}{\pi(\hbar v_F)^2L}\left|E\right|,
\label{eq_DOS}
\end{eqnarray}
where a factor of $4$ arises from the spin and node degeneracy at each of the WSM semi-infinite contacts.
With this consideration, the expression for the node component of the current $I_{\xi}$ becomes
\begin{eqnarray}
I_{\xi} = e v_F \sum_{n,\lambda}  \mathcal{T}(E_{\lambda}^{\xi}(n))\left[ f_L(E_{\lambda}^{\xi}(n)) - f_R(E_{\lambda}^{\xi}(n))\right]
\label{eq_current}
\end{eqnarray}
with the total current given by $I = I_{+} + I_{-}$, and the effective transmission function
\begin{eqnarray}
\mathcal{T}(E) \equiv \frac{8}{\pi^2}\sum_{m,p}\frac{(-1)^{p+1}}{\tilde{\sigma}(E)(4p^2-1)}
e^{i(\delta_m - \delta_{m-2p})}\sin\delta_m\sin\delta_{m-2p}
\end{eqnarray}
The electrical conductance is obtained, from definition, as the voltage derivative of the total current $I = I_{+} + I_{-}$ at constant temperature,
\begin{eqnarray}
G(T,V) &=& \left.\frac{\partial I}{\partial V}\right|_{T}\nonumber\\
&=& \frac{e^2 v_F}{k_B T} \sum_{\lambda,n,\xi}\mathcal{T}(E_{\lambda}^{\xi}(n))f_L(E_{\lambda}^{\xi}(n))\left[
1 - f_L(E_{\lambda}^{\xi}(n))\right], 
\label{eq_conductance}
\end{eqnarray}
where we assumed the conditons $\mu_L = \mu_R + eV$, $T_L = T_R = T$, and we applied the identity
\begin{eqnarray}
\frac{\partial f_L (E)}{\partial V} &=& \frac{e}{k_B T }\left[1 - f_L(E) \right] f_L(E).
\end{eqnarray}

\section{Heat current and thermoelectric transport coefficients}
\label{sec:thermo}

Under similar arguments as explained above, within the Landauer picture\cite{Datta_1,Datta_2,Nazarov} each mode in the energy spectrum arising from the node $K_{\xi}$ carries an energy flow rate
\begin{eqnarray}
\delta \dot{U}_{\xi} = W\,L v_F \,\left\{D_{L}(E) f_{L}(E) \,E  - D_{R}(E)f_{R}(E)\,E   \right\}\bar{T}_{\xi}(E)dE
\end{eqnarray}
Hence, according to the Thermodynamics relation $T dS = dU - \mu dN$, the net heat current transmitted across the junction arising from the node at $K_{\xi}$ is
\begin{eqnarray}
\dot{Q}_{\xi} &=& \dot{U}_{\xi} -\left(\mu_L - \mu_R\right) I_{\xi}\nonumber\\
 &=& W\,L v_F\,\int_{-\infty}^{\infty}\left\{D_{L}(E) f_{L}(E) \, E  - D_{R}(E)f_{R}(E)\,  E \right\}\bar{T}_{\xi}(E)dE
- (\mu_L - \mu_R) I_{\xi}.
\label{eq_heatcurrent}
\end{eqnarray}
Here, the Fermi distribution functions are evaluated at the local temperature and chemical potential of each contact, i.e. $f_L(E) \equiv \left(\exp[(E-\mu_L)/(k_B T_L)]+1\right)^{-1}$
and $f_R(E) \equiv \left(\exp[(E-\mu_R)/(k_B T_R)]+1\right)^{-1}$, respectively.

By considering the expression Eq.(\ref{eq_DOS}) for the density of states at each semi-infinite contact region, we obtain by integrating Eq.(\ref{eq_heatcurrent}) explicitly
\begin{eqnarray}
\dot{U}_{\xi} =  v_F \sum_{n,\lambda}  \mathcal{T}(E_{\lambda}^{\xi}(n))\left[ (E_{\lambda}^{\xi}(n) - \mu_L)f_L(E_{\lambda}^{\xi}(n)) - (E_{\lambda}^{\xi}(n) - \mu_R)f_R(E_{\lambda}^{\xi}(n))\right]
\end{eqnarray}
with the total heat flux given by the superposition from both Weyl nodes $\dot{U} = \dot{U}_{+} + \dot{U}_{-}$.

The thermal conductance is defined, as usual, under conditions where there is no net electric current flowing through the junction ($I=0$)
\begin{eqnarray}
\kappa(T,V) = - \left.\frac{\partial \dot{Q}}{\partial \Delta T}\right|_{I = 0} = - \left.\frac{\partial \dot{U}}{\partial \Delta T}\right|_{I = 0}
\label{eq_kappa}
\end{eqnarray}
with $\Delta T = T_R - T_L$ the temperature difference between the contacts. We remark that this analysis only involves the electronic contribution
to the heat conduction. At finite temperatures, the onset of lattice vibrations will induce a phonon contribution to the total thermal conductance \cite{Ziman,Madelung,Munoz_010,Munoz_CRC_016}.
Moreover, at temperatures on the order of the Debye or Bloch-Gr\"uneisen \cite{Ziman,Madelung,Munoz_012,Munoz_CRC_016} temperatures of the material, the presence of phonon modes will limit
thermal conduction due to electron-phonon scattering\cite{Ziman,Munoz_012}. Therefore, at these higher temperatures the Landauer picture of transport breaks down and
other theoretical approaches are more appropriate, such as Green's functions \cite{Kirchner_12,Munoz_013} or the semiclassical Boltzmann equation \cite{Ziman,Madelung,Munoz_012,Munoz_CRC_016}.

The condition of vanishing electric current defines an implicit relation between the voltage difference and the thermal gradient across the junction,
by $I(\Delta T, V,T) = 0$. This implicit relation allows us to express the Seebeck coefficient by using the implicit function theorem
\begin{eqnarray}
S(T,V) = -\left.\frac{\partial V}{\partial \Delta T}\right|_{I=0,T}  =\frac{\displaystyle \left.\frac{\partial I}{\partial \Delta T}\right|_{T,V}}{\displaystyle \left.\frac{\partial I}{\partial V}\right|_{T,\Delta T}}
\label{eq_Seebeck}
\end{eqnarray}
where $\Delta T(V,T)$ is obtained as the solution of the equation $I(T,V,\Delta T) = 0$.

Taking into account that $T_L = T$, $T_R = T + \Delta T$, $\mu_R = \mu$ and $\mu_L = \mu + e V$, we have
\begin{eqnarray}
f_L(E) &=& f(E- \mu -e V,T) = \left(\exp((E-\mu-e V)/(k_B T)) + 1\right)^{-1}\nonumber\\
f_R(E) &=& f(E-\mu, T + \Delta T) = \left(\exp((E - \mu)/(k_B(T + \Delta T)) + 1\right)^{-1},
\end{eqnarray}
and hence the derivatives
\begin{eqnarray}
\frac{\partial f_R (E)}{\partial \Delta T} &=& \frac{E - \mu}{k_B\left(T + \Delta T\right)^2}\left[1 - f_R(E) \right] f_R(E)\nonumber\\
\frac{\partial f_L (E)}{\partial V} &=& \frac{e}{k_B T }\left[1 - f_L(E) \right] f_L(E)
\end{eqnarray}
we obtain the explicit analytical expression that results from Eq.(\ref{eq_Seebeck})
\begin{eqnarray}
S(T,V) =-\frac{T\sum_{\lambda,n,\xi}\mathcal{T}(E_{\lambda}^{\xi}(n))\left(E_{\lambda}^{\xi}(n) - \mu\right) f_R(E_{\lambda}^{\xi}(n))\left[1- f_R(E_{\lambda}^{\xi}(n)) \right]}
{e (T + \Delta T)^2\sum_{\lambda,n,\xi}\mathcal{T}(E_{\lambda}^{\xi}(n))f_L(E_{\lambda}^{\xi}(n))\left[1- f_L(E_{\lambda}^{\xi}(n)) \right]}
\label{eq_Seebeck_formula}
\end{eqnarray}

Following the argument above, the thermal conductance defined in Eq.(\ref{eq_kappa}) is calculated by the chain rule
\begin{eqnarray}
\kappa(T,V) &=& -\left.\frac{\partial U}{\partial\Delta T}\right|_{T,V} -\left.\frac{\partial V}{\partial  \Delta T}\right|_{I=0,V}\left.\frac{\partial U}{\partial V}\right|_{ T,\Delta T}\nonumber\\
&=& -\left.\frac{\partial U}{\partial\Delta T}\right|_{T,V} + S(T,V)\left.\frac{\partial U}{\partial V}\right|_{ T,\Delta T}
\end{eqnarray}

The explicit analytical expression obtained from this formula is
\begin{eqnarray}
\kappa(T,V) &=&  \frac{v_F }{k_B (T + \Delta T)^2}\sum_{\lambda,n,\xi}\mathcal{T}(E_{\lambda}^{\xi}(n)) E_{\lambda}^{\xi}(n) \left[ E_{\lambda}^{\xi}(n) - \mu  \right] f_R(E_{\lambda}^{\xi}(n))\left[
1 - f_R(E_{\lambda}^{\xi}(n))\right]\\
&+&S(T,V)\frac{e v_F}{k_B T} \sum_{\lambda,n,\xi}\mathcal{T}(E_{\lambda}^{\xi}(n))\left[E_{\lambda}^{\xi}(n) - \mu\right]f_L(E_{\lambda}^{\xi}(n))\left[
1 - f_L(E_{\lambda}^{\xi}(n))\right]. \nonumber
\label{eq_kappa_formula}
\end{eqnarray}



\section{Results and Discussion}

\begin{figure}[hbt]
\centering
\subfigure[]{\includegraphics[width=0.483\textwidth]{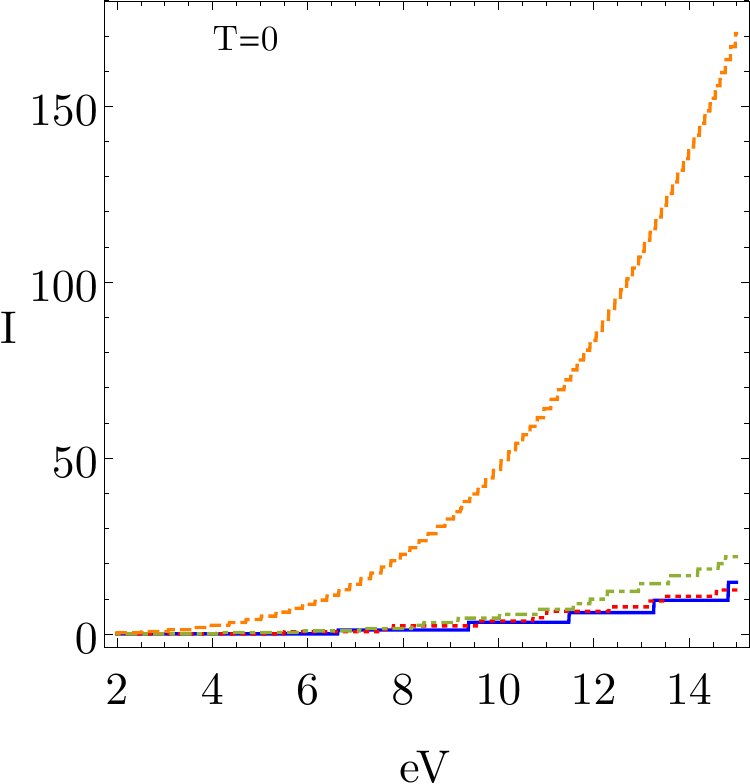}\label{fig2a} }
\hskip .1cm 
\subfigure[]{\includegraphics[width=0.47\textwidth]{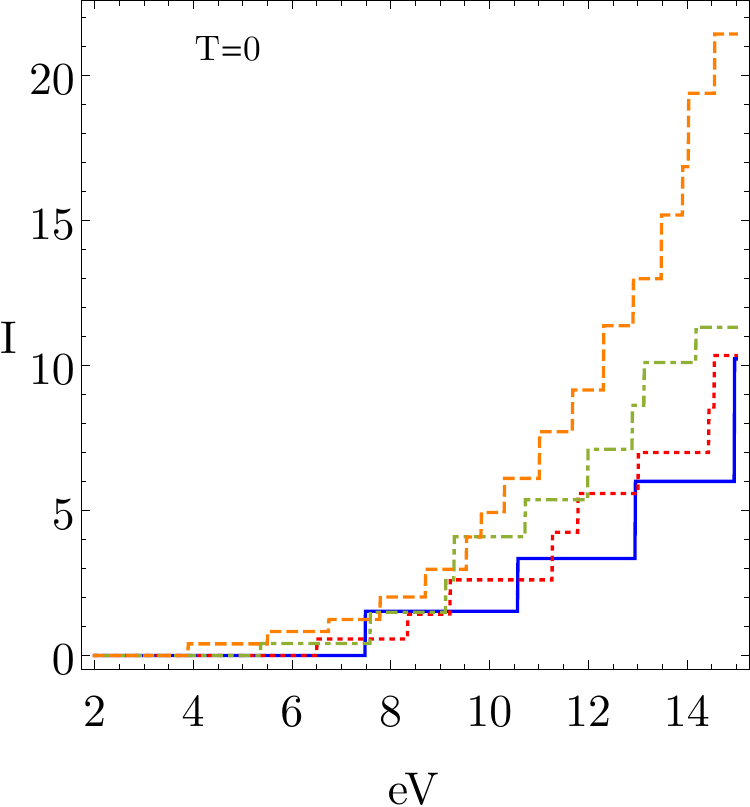}\label{fig2b}}
\caption{(Color online) Current ({\bf in units of $e v_F/a$}) calculated from the analytical Eq.(\ref{eq_current}), as a function of applied bias $V$ ({\bf in units of $\hbar v_F/(a\,e)$}), for fixed $B_0a^2=22\tilde{\phi}_0$ and different values of the torsion angle $\theta$. The solid (blue) line corresponds to $\theta=0$ ($B_Sa^2=0$), the dotted (red) line corresponds to $\theta=5^\circ$ ($B_Sa^2=6.8\tilde{\phi}_0$), the dotdashed (green) line corresponds to $\theta=10^\circ$ ($B_Sa^2=13.6\tilde{\phi}_0$) and the dashed (orange) line corresponds to $\theta=15^\circ$($B_Sa^2=20.4\tilde{\phi}_0$), with $\tilde{\phi}_{0}\equiv \hbar v_F /e$. The subfigure (b) represents the same set of torsional angles, submitted to a magnetic field $B_0a^2=28\tilde{\phi}_0$.}
\label{fig2}
\end{figure}

\begin{figure}[hbt]
\centering
\subfigure[]{\includegraphics[width=0.45\textwidth]{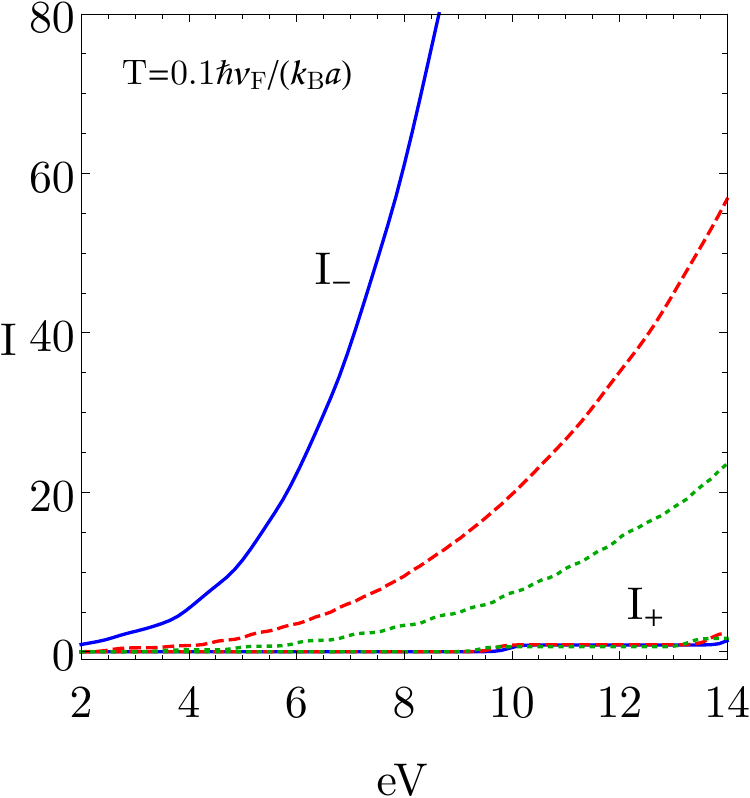}\label{fig3a}}
\subfigure[]{\includegraphics[width=0.45\textwidth]{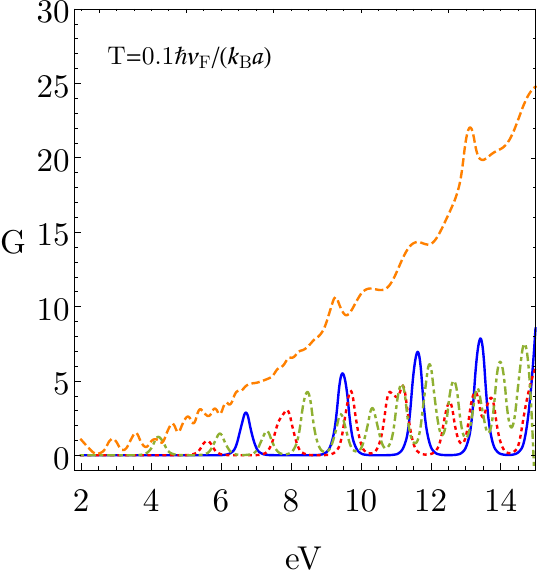}\label{fig3b}}
\caption{(Color online)(a) Node-polarized components of the current ({\bf in units of $e v_F/a$}), calculated from the analytical Eq.(\ref{eq_current}), as a function of the applied bias $e V$ ({\bf in units of $\hbar v_F/a$}), at finite temperature $T_L=T_R=T=0.1\,\hbar v_F/ (k_B a)$, fixed $B_0a^2=25\tilde{\phi}_0$ and different  values of the torsion angle $\theta$.
The solid (blue) lines correspond to $\theta=18^\circ$ ($B_Sa^2=24.5\tilde{\phi}_0$), the dashed (red) lines correspond to  $\theta=16^\circ$ ($B_Sa^2=21.8\tilde{\phi}_0$) and  the dotted (green) lines correspond to $\theta=14^\circ$ ($B_Sa^2=19\tilde{\phi}_0$), with $\tilde{\phi}_{0}\equiv \hbar v_F/e$. (b) Conductance (in units of $e^2/\hbar$) as a function of bias $e V$ (in units of $\hbar v_F/a$), calculated as the voltage-derivative of the analytical Eq.(\ref{eq_current}), for fixed $B_0a^2=22.5\tilde{\phi}_0$ and different values of the torsion angle $\theta$. The solid (blue) line corresponds to $\theta=0$, the dotted (red) line corresponds to $\theta=5^\circ$, the dotdashed (green) line corresponds to $\theta=10^\circ$ and the dashed (orange) line corresponds to $\theta=15^\circ$, with $\tilde{\phi}_{0}\equiv \hbar v_F /e$. }
\label{fig3}
\end{figure}

In the following section, we will present results of the electronic and thermal transport properties for different values of the parameters of the model (external magnetic field, torsional strain, temperature, etc. ) 

To approximate the magnitude of the strain-induced pseudo-magnetic field, we follow Pikulin and collaborators \cite{Pikulin-2016} that estimated $B_S\approx1.8\times 10^{-3}T$ per angular degree of torsion for a cylinder of length $L\sim 100$ nm and radius $a\sim 50$ nm. In addition, we define the scaled flux quantum $\displaystyle\tilde{\phi}_{0}\equiv \frac{\hbar v_F}{e}=\frac{1}{2\pi}\frac{v_F}{c}\frac{hc}{e}=\frac{1}{2\pi}\frac{1.5}{300}\cdot4.14\times10^5$ T$\mathring{\text{A}}^2\approx330$ T$\mathring{\text{A}}^2$. From these values we obtain:
\begin{equation}
 B_Sa^2=1.36\theta \tilde{\phi}_{0},
 \label{eq:torsionangle}
\end{equation}
where $\theta$ is the torsion (twist) angle in degrees. In analogy to the case of graphene discussed in \cite{Munoz-2017}, we present here the results for the total current $I = I_{+} + I_{-}$ (in units of $ev_F/a$) from Eq.~\eqref{eq_current} and conductance for different values of the torsion angle $\theta$ and different values of the external magnetic field.

In Fig.\ref{fig2a} we show the total current for $T_L=T_R=T = 0$, as a function of the applied bias voltage $V$, for a fixed value of the external magnetic field $B_0a^2 =22\tilde{\phi}_{0}$. The different curves correspond to different values of the torsion angle $0 \le \theta \le 15^o$, which correspond to different values for the pseudo-magnetic field $B_S$. An important feature of the current-voltage curves are the plateaus, which are more distinguishable at lower values of torsional strain. This effect can be understood from the elastic scattering (see Eq.\eqref{eq_transm}) assumption, that imposes the constraint that for the incident particle to be transmitted across the strained scattering region, its energy must be resonant with one of the pseudo-Landau level eigenstates. 
This condition is always fulfilled thanks to the quasi-continuum distribution of energy values at the contacts, for an interval within the energy window imposed by the external bias voltage. At low but finite temperature, the plateaus are still present, but they are smoother than at zero temperature and they eventually smear away at high enough temperatures (see Ref. \cite{Soto_018}), as can be appreciated in Fig \ref{fig3a}.

The effect of the external magnetic field $B_0$ on the current-voltage curves is compared in Fig.\ref{fig2a} and Fig.\ref{fig2b}, respectively. At zero and finite temperatures, one can observe that for the same values of torsion angle, i.e. $\theta = (5^\circ,10^\circ,15^\circ)$, the total current decreases as the external magnetic field is increased from $B_{0}a^2 = 22\tilde{\phi}_0$ (in Fig.\ref{fig2a}) towards $B_{0}a^2 = 28\tilde{\phi}_0$ (in Fig.\ref{fig2b}). The origin of this effect is at the density of the pseudo-Landau level spectrum at the nodes. For a fixed value of $\theta$ and hence of $B_S$, the effective pseudo-magnetic field $|B_{\xi}| = |B_{0} + \xi B_S|$ increases as $B_0$ increases (we are taking $B_0>B_S$ in Figures \ref{fig2a},\ref{fig2b}). This reduces the spectral density at both nodes given that $E_n^{(\xi)} \sim \sqrt{|B_{\xi}|n}$, thus providing fewer channels for the incident particles to be transmitted and hence decreasing the total current $I$ across the contacts.

Another remarkable feature of the current-voltage curves is the node polarization effect, due to a relative enhancement of the node-polarized contribution $I_{-}$ arising from the $\mathbf{K}_{-}$ node, as can be clearly observed in Fig.\ref{fig3a}, where the two nodes components $I_{+}$ and $I_{-}$ of the total current are represented at finite temperature. This effect is even stronger when $B_S$ is closer to $B_0$ and can be understood by
noticing that the effective pseudo-magnetic field $|B_{-}| = |B_{0} - B_S|$ decreases as $B_S$ increases, thus leading to a relatively higher spectral density associated to the $\mathbf{K}_{-}$-node where $E_n^{(-)} \sim \sqrt{|B_{-}|n}$. Therefore, more channels are available for the incident particles to be transmitted per finite interval of bias voltage imposed at the contacts, thus increasing the current $I_{-}$ associated to the $\mathbf{K}_{-}$-node. This effect can also be appreciated at finite temperatures, as
seen in Fig.~\ref{fig3a}, and it could be used to design a node-sensitive filter as discussed in Ref. \cite{Soto_018}.

The differential conductance $G(V,T) = dI/dV$ (in units of $e^2/\hbar$) at 
finite temperature $T = 0.1 \,\hbar v_F/ (k_B a)$ is displayed in Fig.~\ref{fig3b} for different values of  the torsion angle $\theta$ and fixed $B_0=22.5\tilde{\phi}_0/a^2$. A characteristic trend of oscillations is observed, which are consistent with the staircase behavior of the current observed in Figs.~\ref{fig2a},\ref{fig2b}.

\subsection{Thermal transport examples}

Let us now discuss the behaviour of the thermal transport coefficients, whose analytical formulas where obtained in section \ref{sec:thermo}, under the light of a few explicit examples.

\begin{figure}[t!]
\centering
\subfigure[]{\includegraphics[width=0.495\textwidth]{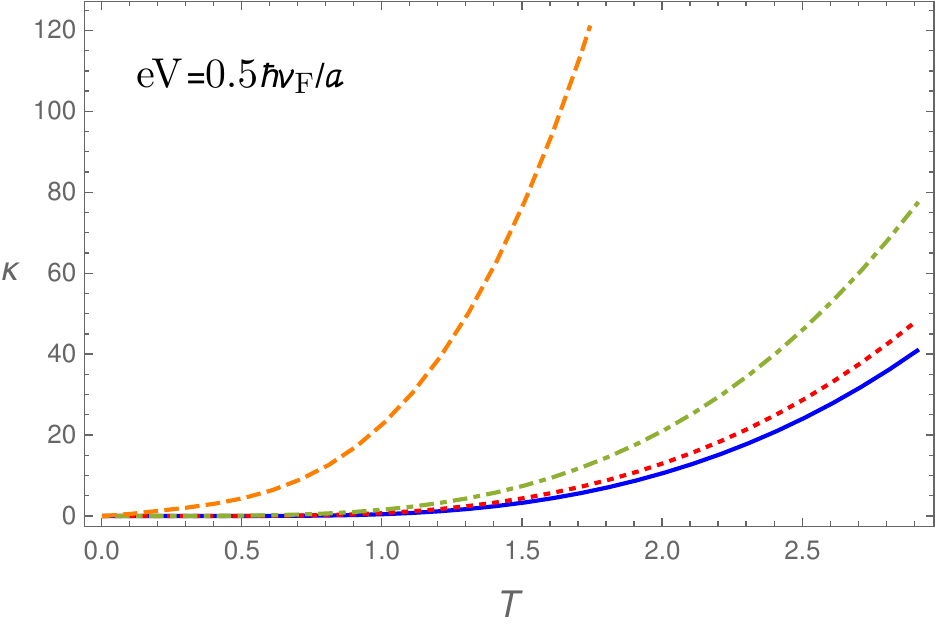}\label{fig4a}}
\subfigure[]{\includegraphics[width=0.495\textwidth]{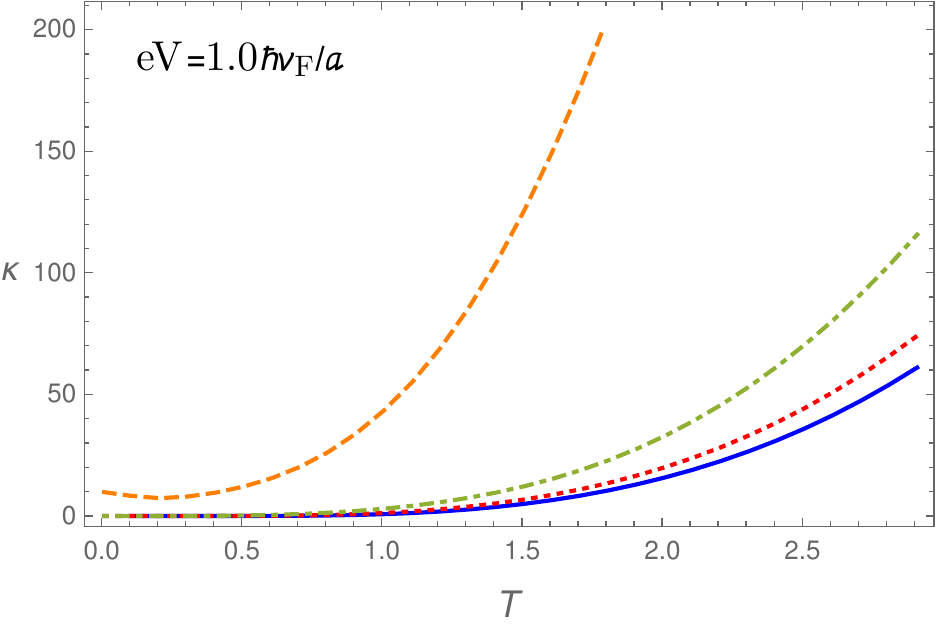}\label{fig4b}}
\caption{(Color online) Thermal conductance ({\bf in units of $k_B v_F/a$}) as a function of temperature $T$ ({\bf in units of $\hbar\, v_F/(a k_B)$}), calculated from the analytical Eq.(\ref{eq_kappa_formula}), for fixed $B_0a^2=22\tilde{\phi}_0$  and different values of the torsion angle $\theta$. The solid (blue) line corresponds to $\theta=0$, the dotted (red) line corresponds to $\theta=5^\circ$, the dotdashed (green) line corresponds to $\theta=10^\circ$ and the dashed (orange) line corresponds to $\theta=15^\circ$, with $\tilde{\phi}_{0}\equiv \hbar v_F /e$. (a) Thermal conductance for a bias of $e V = 0.5\, \hbar v_F/a$ and (b) thermal conductance for a bias of $e V = 1.0 \,\hbar v_F/ a$}
\label{fig4}
\end{figure}

The thermal conductance (in units of $k_B v_F/a$) through the strained cylindrical region is represented in Fig.(\ref{fig4}) as a function of temperature, for different magnitudes of strain (represented by its torsional angle $\theta$) and at a constant external magnetic field of $B_0a^2=22\tilde{\phi}_0$. We notice that the thermal conductance increases when increasing mechanical strain, in agreement with the features already discussed for the electrical conductance. This trend is natural to expect, since the same carriers (i.e. Weyl fermions) are involved in the
transport of charge and energy in the material. By comparing the effect of the imposed bias, which is $V = 0.5\,\hbar v_F/(a e)$ in Fig.(\ref{fig4a}) and  $V = 1.0\,\hbar v_F/(a e)$ in Fig.(\ref{fig4b}), it is clear that it has an enhancement effect over the thermal transport. While usually thermal conductivities for bulk materials are typically reported in the linear response regime, in nanoscale junctions and heterostructures it is interesting to study deviations from it. Here, we observe that a combination between the external bias and torsional strain could be used to engineer and control the thermal conductance across this type of structure.  We remark, as discussed at the beginning of Section V, that our model
describes only the electronic contribution to the thermal conductance, with the consequent monotonic behaviour observed in Fig.\ref{fig4}. In experimental conditions, the onset of electron-phonon scattering at higher temperatures
and the presence of impurities and defects in the lattice will limit this value and eventually a decrease with temperature will be observed \cite{Ziman,Madelung,Munoz_CRC_016,Munoz_012}. The treatment of these effects is however beyond the scope of our present analysis, and will be published in a follow-up of this study.

\begin{figure}[t!]
\centering
\subfigure[]{\includegraphics[width=0.495\textwidth]{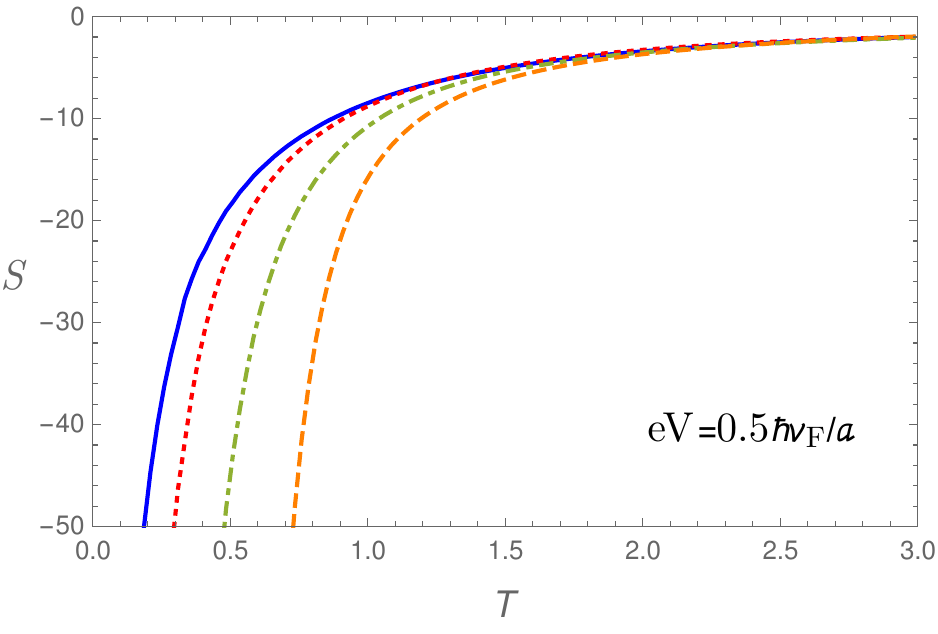}\label{fig5a}}
\subfigure[]{\includegraphics[width=0.495\textwidth]{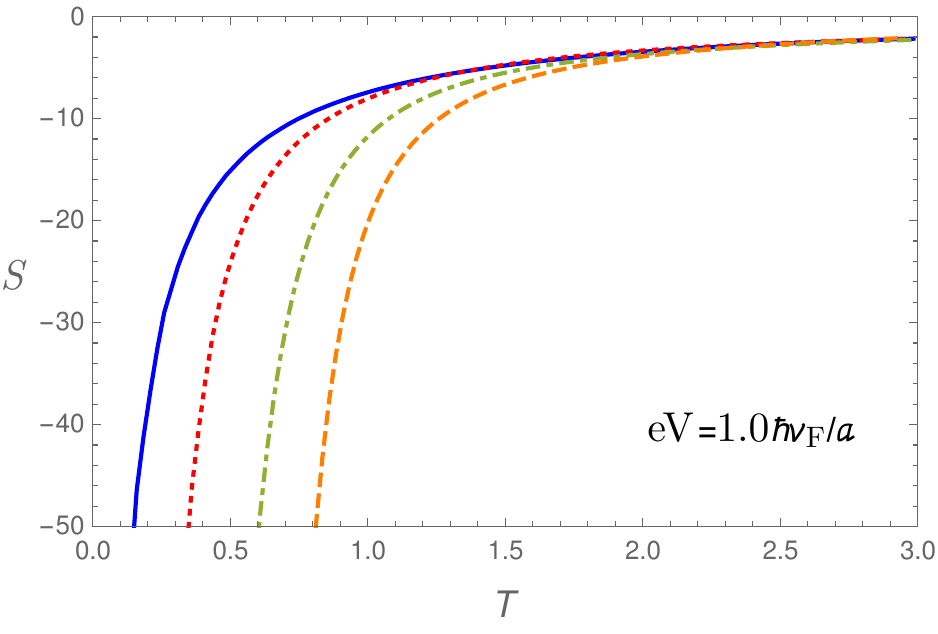}\label{fig5b}}
\caption{(Color online) Seebeck coefficient ({\bf in units of $k_B/e$}) as a function of temperature $T$ ({\bf in units of $\hbar v_F/(a k_B)$}), calculated from the analytical Eq.(\ref{eq_Seebeck_formula}) for fixed $B_0a^2=22\tilde{\phi}_0$ and different values of the torsion angle $\theta$. The solid (blue) line corresponds to $\theta=0$, the dotted (red) line corresponds to $\theta=5^\circ$, the dotdashed (green) line corresponds to $\theta=10^\circ$ and the dashed (orange) line corresponds to $\theta=15^\circ$, with $\tilde{\phi}_{0}\equiv \hbar v_F /e$. (a) Seebeck coefficient for a bias of $e V = 0.5\, \hbar v_F/a$ and (b) Seebeck coefficient for a bias of $e V = 1.0 \,\hbar v_F/ a$}
\label{fig5}
\end{figure}

The calculated Seebeck coefficient (in units of $k_B/e$), according to the analytical formula derived in Eq.(\ref{eq_Seebeck_formula}), is represented for the same examples in Fig.\ref{fig5}. In agreement with the choice of the chemical potential $\mu = 1.0\,\hbar v_F/a > 0$, transport is dominated by negative charge carriers, which translates into a negative Seebeck coefficient as expected. As cleary observed in Figs.\ref{fig5a},\ref{fig5b}, the slope of the Seebeck coefficient a low temperatures is very sensitive to the magnitude of torsional strain, increasing monotonically with the torsional angle. This effect translates into a higher absolute value of the Seebeck coefficient as torsion is applied. By comparing Fig.\ref{fig5a} (with
an applied bias of $V = 0.5\,\hbar v_F/(e\,a)$) with Fig.\ref{fig5b} (at an applied bias of  $V = 1.0\,\hbar v_F/(e\,a)$), we notice that the effect of strain is enhanced by increasing the bias. In summary, we see that the thermopower in this material could in principle be engineered by tuning appropriate combinations of torsional strain and temperature
gradients inducing the desired bias voltages.

\begin{figure}[t!]
\centering
\subfigure[]{\includegraphics[width=0.495\textwidth]{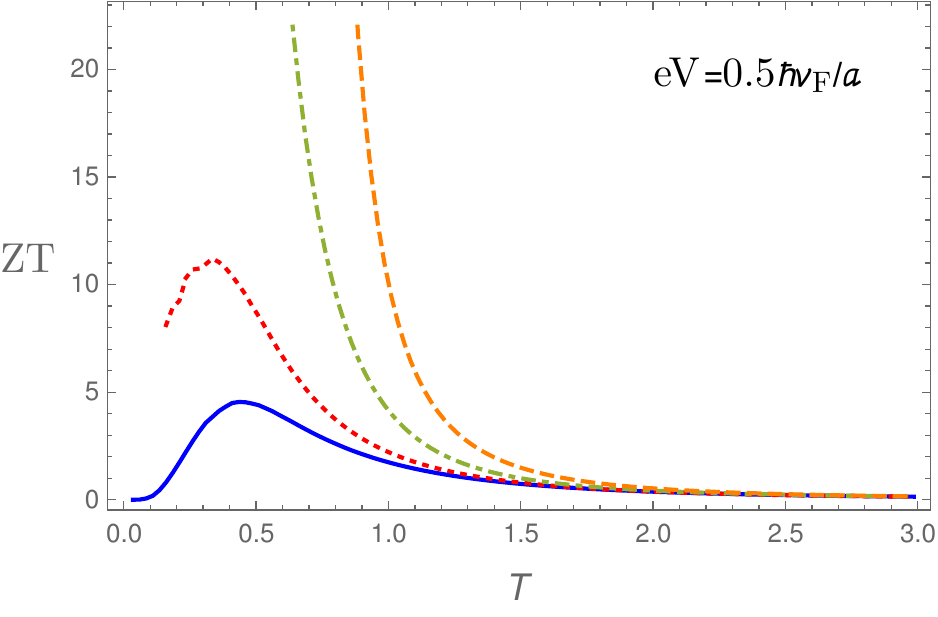}\label{fig5a}}
\subfigure[]{\includegraphics[width=0.495\textwidth]{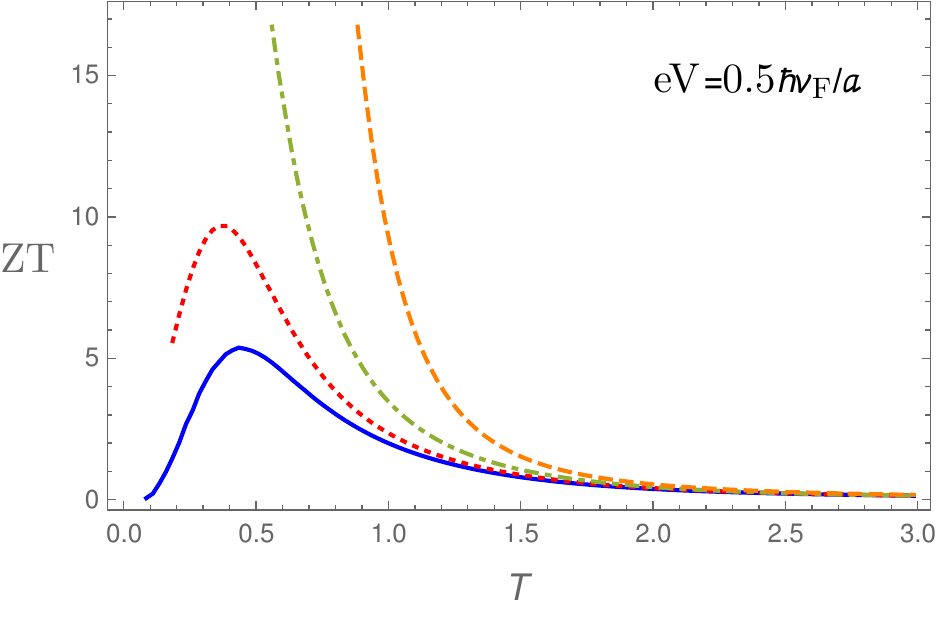}\label{fig5b}}
\caption{(Color online) The figure of merit $ZT$ (dimensionless), calculated from Eq.(\ref{eq_ZT}), for fixed magnetic field $B_0$, at a bias of $e V = 0.5\, \hbar v_F/a$ and different values of the torsion angle $\theta$ is represented as a function of temperature $T$ ({\bf in units of $\hbar v_F/(a k_B)$}). The solid (blue) line corresponds to $\theta=0$, the dotted (red) line corresponds to $\theta=5^\circ$, the dotdashed (green) line corresponds to $\theta=10^\circ$ and the dashed (orange) line corresponds to $\theta=15^\circ$, with $\tilde{\phi}_{0}\equiv \hbar v_F /e$. (a) $B_0a^2=22\tilde{\phi}_0$ and (b) $B_0a^2=28\tilde{\phi}_0$. }
\label{fig6}
\end{figure}

\begin{figure}[t!]
\centering
\subfigure[]{\includegraphics[width=0.495\textwidth]{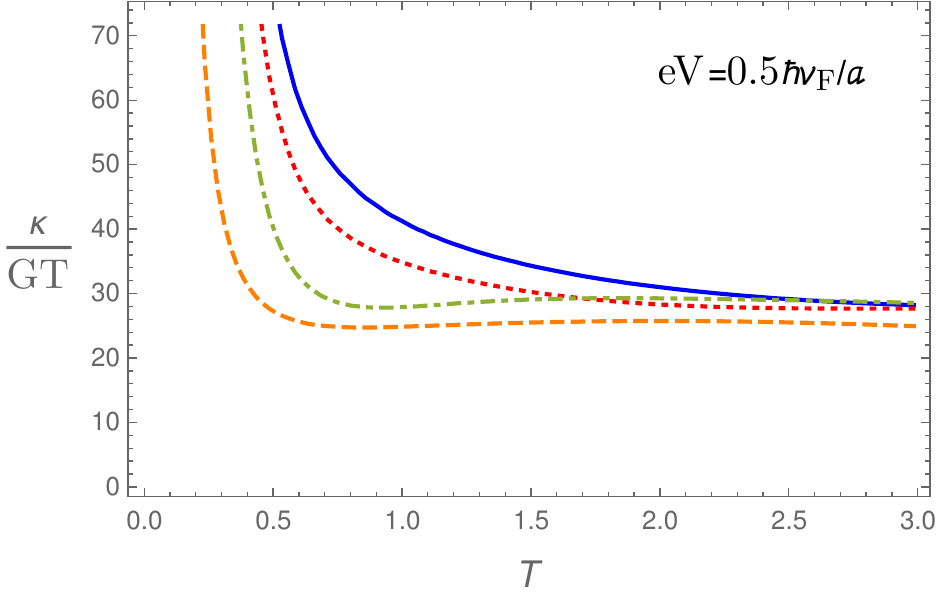}\label{fig5a}}
\subfigure[]{\includegraphics[width=0.495\textwidth]{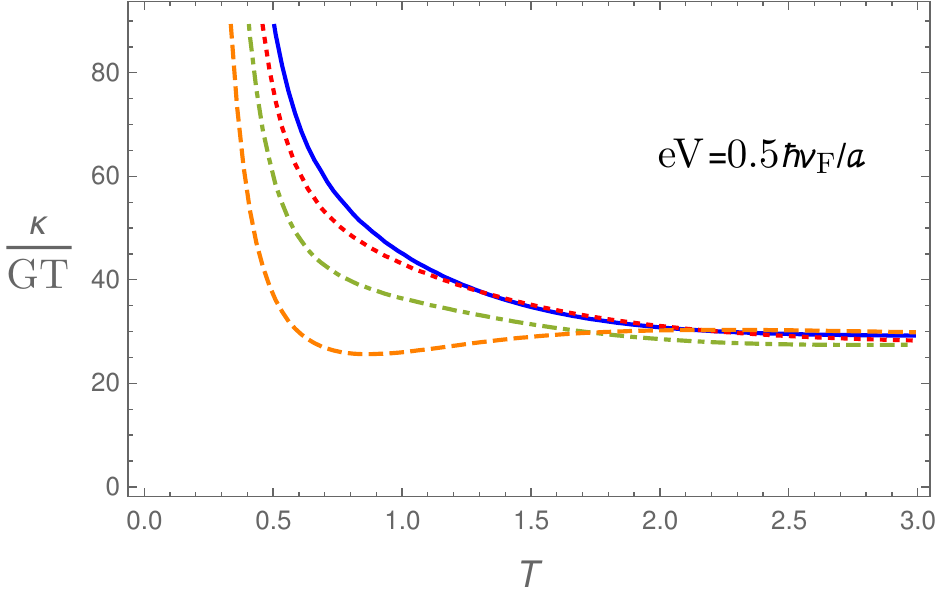}\label{fig5b}}
\caption{(Color online) The Lorenz number $L$ ({\bf{in units of $k_B^2/e^2$} }) calculated from Eq.(\ref{eq_Lorenz}), for fixed magnetic field $B_0$, at a bias of $e V = 0.5\, \hbar v_F/a$ and different values of the torsion angle $\theta$, is represented as a function of temperature $T$ ({\bf in units of $\hbar v_F/(a k_B)$}). The solid (blue) line corresponds to $\theta=0$, the dotted (red) line corresponds to $\theta=5^\circ$, the dotdashed (green) line corresponds to $\theta=10^\circ$ and the dashed (orange) line corresponds to $\theta=15^\circ$, with $\tilde{\phi}_{0}\equiv \hbar v_F /e$. (a) $B_0a^2=22\tilde{\phi}_0$ and (b) $B_0a^2=28\tilde{\phi}_0$. }
\label{fig7}
\end{figure}

Finally, in order to characterize the thermoelectric performance of the WSM nanojunction, it is important to
explore the magnitude of the figure of merit $ZT$, defined by the standard formula
\begin{eqnarray}
ZT = S^2  \frac{T\,G(T,V)}{\kappa(T,V)}.
\label{eq_ZT}
\end{eqnarray}
Numerical values of $ZT$, as a function of temperature and strain, are presented in Fig.\ref{fig6} for two different values of the external magnetic field $B_0$. It is remarkable to notice that extremely high values of the figure of merit can be achieved through the combination of magnetic field and torsional strain. Moreover, at fixed magnetic field,
$ZT$ increases with the magnitude of the torsional strain. This is greatly due to the enhancement effect that strain induces over the electronic conductance, as was explained in detail at the beginning of this section,
and the effect is more drastic at low temperatures. As seen in the zero temperature limit of the electronic conductance, Fig.\ref{fig2}, a strong deviation from the metallic behavior is observed
in the staircase pattern that reflects the Landau level spectrum. As temperature increases, however, the electronic conductance curve smears down. It is conceptually
instructive to examine this deviation  from the normal metal behavior by studying the  Lorenz number as a function of temperature for this system, i.e.
\begin{eqnarray}
L = \frac{\kappa(T,V)}{T G(T,V)}.
\label{eq_Lorenz}
\end{eqnarray}
The Lorenz number is represented, at fixed bias and magnetic field, as a function of temperature for different values of torsion in Fig.\ref{fig7}. 
It is observed that for temperatures higher than $T > 2.0\, \hbar v_F/(k_B\,a)$, the Lorenz number tends to a constant value, independent of
temperature and strain, in agreement with the Wiedemann-Franz law, as expected for normal metallic behavior. However, strong
deviations from the Wiedemann-Franz law are observed at low temperatures, in agreement with the non-metallic behavior of
the electronic conductance due to the Landau-level spectrum. It is precisely this effect that allows for extremely high
$ZT$ values at low temperatures, in agreement with experimental evidence recently reported in the literature \cite{Skinner_018} that suggested values as high as $ZT \sim 10$.
{\color{red} One can see from Fig \ref{fig6} that the maximum value for $ZT$ occurs at temperatures near $0.5\hbar v_F/(a k_B)$, which for $a=50$nm and $v_F=1.5\cdot 10^5$ m/s is on the order of $T\sim 10 K$. Although this temperature is quite low for practical applications, it is still interesting to see that such high values of $ZT$ can be reached in principle.}

\section{Conclusions and Summary}

We studied the thermal transport in WSM submitted to torsional strain and an external magnetic field, extending our previous work in the electronic transport properties \cite{Soto_018}. 
Our theoretical model combines scattering theory with the Landauer approach for electronic transport \cite{Soto_018,Munoz-2017}, allowing us to find explicit analytical expressions
for the thermoelectric transport coefficients. From our analytical results, we predict a node-polarization effect on the current \cite{Soto_018} (analogous to the valley polarization observed in graphene \cite{Low_010,Settnes_016}), due to the combination of torsional strain and an external magnetic field. As discussed in detail in Ref.\cite{Soto_018}, this is a consequence of
a node-selective enhancement of the spectral density due to the symmetry breaking between the nodes induced by the linear combination of the strain pseudomagnetic field and the $z$-component of the external magnetic field, that have different relative signs at each node: The physical magnetic field breaks time-reversal symmetry,
whereas the strain field does not. Therefore, the node-filtering effect is robust in different geometries,
and is not only a property of the cylindrical configuration chosen for the explicit analytical solution provided in our work. As discussed in Ref.\cite{Soto_018}, this effect could be used
to develop a strain sensor in WSMs. In addition, we reported quantitative calculations of the thermal conductance and Seebeck coefficient in this configuration. Our theoretical results show that both coefficients are very sensitive to strain, displaying a significant enhancement as the torsional angle is increased. Given that the same carriers, i.e. Weyl fermions,
are responsible for the electric and thermal transport, it is clear that the thermal current (and thus the thermal conductance) increases with torsion by the same mechanism as the electric current. Remarkably, these results suggest that it might be possible to engineer the thermoelectric coefficients (thermal conductance and thermopower) either by imposing
a mechanical (torsional) strain or by appropriately tuning the intensity of an external magnetic field, depending on which of this variables may be easier to control experimentally. 
Remarkably, our theoretical model suggests that under a combination of an external magnetic field
and torsional strain, it might be possible to achieve extremely high thermoelectric performance, characterized by a figure of merit 
$ZT > 10$ at low temperatures ($T < 2.0\, \hbar v_F /(k_B\, a)$). Moreover, this conclusion seems to be supported
by recent experimental evidence \cite{Skinner_018}, thus providing a very promising proof of principle for highly efficient thermoelectric devices.

\section*{Acknowledgements}

This work was in part supported by Fondecyt (Chile) Grant No. 11160542 (R. S.-G.).


\begin{thebibliography}{36}%
\makeatletter
\providecommand \@ifxundefined [1]{%
 \@ifx{#1\undefined}
}%
\providecommand \@ifnum [1]{%
 \ifnum #1\expandafter \@firstoftwo
 \else \expandafter \@secondoftwo
 \fi
}%
\providecommand \@ifx [1]{%
 \ifx #1\expandafter \@firstoftwo
 \else \expandafter \@secondoftwo
 \fi
}%
\providecommand \natexlab [1]{#1}%
\providecommand \enquote  [1]{``#1''}%
\providecommand \bibnamefont  [1]{#1}%
\providecommand \bibfnamefont [1]{#1}%
\providecommand \citenamefont [1]{#1}%
\providecommand \href@noop [0]{\@secondoftwo}%
\providecommand \href [0]{\begingroup \@sanitize@url \@href}%
\providecommand \@href[1]{\@@startlink{#1}\@@href}%
\providecommand \@@href[1]{\endgroup#1\@@endlink}%
\providecommand \@sanitize@url [0]{\catcode `\\12\catcode `\$12\catcode
  `\&12\catcode `\#12\catcode `\^12\catcode `\_12\catcode `\%12\relax}%
\providecommand \@@startlink[1]{}%
\providecommand \@@endlink[0]{}%
\providecommand \url  [0]{\begingroup\@sanitize@url \@url }%
\providecommand \@url [1]{\endgroup\@href {#1}{\urlprefix }}%
\providecommand \urlprefix  [0]{URL }%
\providecommand \Eprint [0]{\href }%
\providecommand \doibase [0]{http://dx.doi.org/}%
\providecommand \selectlanguage [0]{\@gobble}%
\providecommand \bibinfo  [0]{\@secondoftwo}%
\providecommand \bibfield  [0]{\@secondoftwo}%
\providecommand \translation [1]{[#1]}%
\providecommand \BibitemOpen [0]{}%
\providecommand \bibitemStop [0]{}%
\providecommand \bibitemNoStop [0]{.\EOS\space}%
\providecommand \EOS [0]{\spacefactor3000\relax}%
\providecommand \BibitemShut  [1]{\csname bibitem#1\endcsname}%
\let\auto@bib@innerbib\@empty
\bibitem [{\citenamefont {Soto-Garrido}\ and\ \citenamefont
  {Mu{\~n}oz}(2018)}]{Soto_018}%
  \BibitemOpen
  \bibfield  {author} {\bibinfo {author} {\bibfnamefont {R.}~\bibnamefont
  {Soto-Garrido}}\ and\ \bibinfo {author} {\bibfnamefont {E.}~\bibnamefont
  {Mu{\~n}oz}},\ }\href@noop {} {\bibfield  {journal} {\bibinfo  {journal} {J.
  Phys.: Condens. Matter}\ }\textbf {\bibinfo {volume} {30}},\ \bibinfo {pages}
  {195302} (\bibinfo {year} {2018})}\BibitemShut {NoStop}%
\bibitem [{\citenamefont {Armitage}, \citenamefont {Mele},\ and\ \citenamefont
  {Vishwanath}(2018)}]{armitage2017}%
  \BibitemOpen
  \bibfield  {author} {\bibinfo {author} {\bibfnamefont {N.~P.}\ \bibnamefont
  {Armitage}}, \bibinfo {author} {\bibfnamefont {E.~J.}\ \bibnamefont {Mele}},
  \ and\ \bibinfo {author} {\bibfnamefont {A.}~\bibnamefont {Vishwanath}},\
  }\href@noop {} {\bibfield  {journal} {\bibinfo  {journal} {Rev. Mod. Phys.}\
  }\textbf {\bibinfo {volume} {90}},\ \bibinfo {pages} {015001} (\bibinfo
  {year} {2018})}\BibitemShut {NoStop}%
\bibitem [{\citenamefont {Hasan}\ \emph {et~al.}(2017)\citenamefont {Hasan},
  \citenamefont {Xu}, \citenamefont {Beloposky},\ and\ \citenamefont
  {Huang}}]{Hasan_017}%
  \BibitemOpen
  \bibfield  {author} {\bibinfo {author} {\bibfnamefont {M.~Z.}\ \bibnamefont
  {Hasan}}, \bibinfo {author} {\bibfnamefont {S.-Y.}\ \bibnamefont {Xu}},
  \bibinfo {author} {\bibfnamefont {I.}~\bibnamefont {Beloposky}}, \ and\
  \bibinfo {author} {\bibfnamefont {S.-M.}\ \bibnamefont {Huang}},\ }\href@noop
  {} {\bibfield  {journal} {\bibinfo  {journal} {Annu. Rev. Condens. Matter
  Phys.}\ }\textbf {\bibinfo {volume} {8}},\ \bibinfo {pages} {289} (\bibinfo
  {year} {2017})}\BibitemShut {NoStop}%
\bibitem [{\citenamefont {Yan}\ and\ \citenamefont {Felser}(2017)}]{Yan_017}%
  \BibitemOpen
  \bibfield  {author} {\bibinfo {author} {\bibfnamefont {B.}~\bibnamefont
  {Yan}}\ and\ \bibinfo {author} {\bibfnamefont {C.}~\bibnamefont {Felser}},\
  }\href@noop {} {\bibfield  {journal} {\bibinfo  {journal} {Annu. Rev.
  Condens. Matter Phys.}\ }\textbf {\bibinfo {volume} {8}},\ \bibinfo {pages}
  {337} (\bibinfo {year} {2017})}\BibitemShut {NoStop}%
\bibitem [{\citenamefont {Burkov}(2018)}]{Burkov_016}%
  \BibitemOpen
  \bibfield  {author} {\bibinfo {author} {\bibfnamefont {A.}~\bibnamefont
  {Burkov}},\ }\href@noop {} {\bibfield  {journal} {\bibinfo  {journal} {Annu.
  Rev. Condens. Matter Phys.}\ }\textbf {\bibinfo {volume} {9}},\ \bibinfo
  {pages} {359} (\bibinfo {year} {2018})}\BibitemShut {NoStop}%
\bibitem [{\citenamefont {Weyl}(1929)}]{Weyl_29}%
  \BibitemOpen
  \bibfield  {author} {\bibinfo {author} {\bibfnamefont {H.}~\bibnamefont
  {Weyl}},\ }\href@noop {} {\bibfield  {journal} {\bibinfo  {journal} {Z.
  Phys.}\ }\textbf {\bibinfo {volume} {56}},\ \bibinfo {pages} {330} (\bibinfo
  {year} {1929})}\BibitemShut {NoStop}%
\bibitem [{\citenamefont {Huang}\ \emph {et~al.}(2016)\citenamefont {Huang},
  \citenamefont {McCormick}, \citenamefont {Ochi}, \citenamefont {Zhao},
  \citenamefont {Suzuki}, \citenamefont {Arita}, \citenamefont {Wu},
  \citenamefont {Mou}, \citenamefont {Cao}, \citenamefont {Yan}, \citenamefont
  {Trivedi},\ and\ \citenamefont {Kaminski}}]{Huang_016}%
  \BibitemOpen
  \bibfield  {author} {\bibinfo {author} {\bibfnamefont {L.}~\bibnamefont
  {Huang}}, \bibinfo {author} {\bibfnamefont {T.~M.}\ \bibnamefont
  {McCormick}}, \bibinfo {author} {\bibfnamefont {M.}~\bibnamefont {Ochi}},
  \bibinfo {author} {\bibfnamefont {Z.}~\bibnamefont {Zhao}}, \bibinfo {author}
  {\bibfnamefont {M.-T.}\ \bibnamefont {Suzuki}}, \bibinfo {author}
  {\bibfnamefont {R.}~\bibnamefont {Arita}}, \bibinfo {author} {\bibfnamefont
  {Y.}~\bibnamefont {Wu}}, \bibinfo {author} {\bibfnamefont {D.}~\bibnamefont
  {Mou}}, \bibinfo {author} {\bibfnamefont {H.}~\bibnamefont {Cao}}, \bibinfo
  {author} {\bibfnamefont {J.}~\bibnamefont {Yan}}, \bibinfo {author}
  {\bibfnamefont {N.}~\bibnamefont {Trivedi}}, \ and\ \bibinfo {author}
  {\bibfnamefont {A.}~\bibnamefont {Kaminski}},\ }\href@noop {} {\bibfield
  {journal} {\bibinfo  {journal} {Nat. Mater.}\ }\textbf {\bibinfo {volume}
  {15}},\ \bibinfo {pages} {1155} (\bibinfo {year} {2016})}\BibitemShut
  {NoStop}%
\bibitem [{\citenamefont {Morimoto}\ and\ \citenamefont
  {Nagaosa}(2016)}]{Morimoto_016}%
  \BibitemOpen
  \bibfield  {author} {\bibinfo {author} {\bibfnamefont {T.}~\bibnamefont
  {Morimoto}}\ and\ \bibinfo {author} {\bibfnamefont {N.}~\bibnamefont
  {Nagaosa}},\ }\href@noop {} {\bibfield  {journal} {\bibinfo  {journal} {Phys.
  Rev. Lett.}\ }\textbf {\bibinfo {volume} {117}},\ \bibinfo {pages} {146603}
  (\bibinfo {year} {2016})}\BibitemShut {NoStop}%
\bibitem [{\citenamefont {Amorim}\ \emph {et~al.}(2016)\citenamefont {Amorim},
  \citenamefont {Cortijo}, \citenamefont {de~Juan}, \citenamefont {Grushin},
  \citenamefont {Guinea}, \citenamefont {Guti\'errez-Rubio}, \citenamefont
  {Ochoa}, \citenamefont {Parente}, \citenamefont {Rold\'an}, \citenamefont
  {San-Jos\'e}, \citenamefont {Schiefele}, \citenamefont {Sturla},\ and\
  \citenamefont {Vozmediano}}]{Amorim_016}%
  \BibitemOpen
  \bibfield  {author} {\bibinfo {author} {\bibfnamefont {B.}~\bibnamefont
  {Amorim}}, \bibinfo {author} {\bibfnamefont {A.}~\bibnamefont {Cortijo}},
  \bibinfo {author} {\bibfnamefont {F.}~\bibnamefont {de~Juan}}, \bibinfo
  {author} {\bibfnamefont {A.~G.}\ \bibnamefont {Grushin}}, \bibinfo {author}
  {\bibfnamefont {F.}~\bibnamefont {Guinea}}, \bibinfo {author} {\bibfnamefont
  {A.}~\bibnamefont {Guti\'errez-Rubio}}, \bibinfo {author} {\bibfnamefont
  {H.}~\bibnamefont {Ochoa}}, \bibinfo {author} {\bibfnamefont
  {V.}~\bibnamefont {Parente}}, \bibinfo {author} {\bibfnamefont
  {R.}~\bibnamefont {Rold\'an}}, \bibinfo {author} {\bibfnamefont
  {P.}~\bibnamefont {San-Jos\'e}}, \bibinfo {author} {\bibfnamefont
  {J.}~\bibnamefont {Schiefele}}, \bibinfo {author} {\bibfnamefont
  {M.}~\bibnamefont {Sturla}}, \ and\ \bibinfo {author} {\bibfnamefont
  {M.~A.~H.}\ \bibnamefont {Vozmediano}},\ }\href@noop {} {\bibfield  {journal}
  {\bibinfo  {journal} {Phys. Rep.}\ }\textbf {\bibinfo {volume} {617}},\
  \bibinfo {pages} {1} (\bibinfo {year} {2016})}\BibitemShut {NoStop}%
\bibitem [{\citenamefont {Naumis}\ \emph {et~al.}(2017)\citenamefont {Naumis},
  \citenamefont {Barraza-Lopez}, \citenamefont {Oliva-Leyva},\ and\
  \citenamefont {Terrones}}]{Naumis-2017}%
  \BibitemOpen
  \bibfield  {author} {\bibinfo {author} {\bibfnamefont {G.~G.}\ \bibnamefont
  {Naumis}}, \bibinfo {author} {\bibfnamefont {S.}~\bibnamefont
  {Barraza-Lopez}}, \bibinfo {author} {\bibfnamefont {M.}~\bibnamefont
  {Oliva-Leyva}}, \ and\ \bibinfo {author} {\bibfnamefont {H.}~\bibnamefont
  {Terrones}},\ }\href@noop {} {\bibfield  {journal} {\bibinfo  {journal} {Rep.
  Prog. Phys.}\ }\textbf {\bibinfo {volume} {80}},\ \bibinfo {pages} {096501}
  (\bibinfo {year} {2017})}\BibitemShut {NoStop}%
\bibitem [{\citenamefont {Cortijo}\ \emph {et~al.}(2015)\citenamefont
  {Cortijo}, \citenamefont {Ferreir\'os}, \citenamefont {Landsteiner},\ and\
  \citenamefont {Vozmediano}}]{Cortijo_015}%
  \BibitemOpen
  \bibfield  {author} {\bibinfo {author} {\bibfnamefont {A.}~\bibnamefont
  {Cortijo}}, \bibinfo {author} {\bibfnamefont {Y.}~\bibnamefont
  {Ferreir\'os}}, \bibinfo {author} {\bibfnamefont {K.}~\bibnamefont
  {Landsteiner}}, \ and\ \bibinfo {author} {\bibfnamefont {M.~A.~H.}\
  \bibnamefont {Vozmediano}},\ }\href@noop {} {\bibfield  {journal} {\bibinfo
  {journal} {Phys. Rev. Lett.}\ }\textbf {\bibinfo {volume} {115}},\ \bibinfo
  {pages} {177202} (\bibinfo {year} {2015})}\BibitemShut {NoStop}%
\bibitem [{\citenamefont {Cortijo}\ \emph
  {et~al.}(2016{\natexlab{a}})\citenamefont {Cortijo}, \citenamefont
  {Kharzeev}, \citenamefont {Landsteiner},\ and\ \citenamefont
  {Vozmediano}}]{Cortijo-2016}%
  \BibitemOpen
  \bibfield  {author} {\bibinfo {author} {\bibfnamefont {A.}~\bibnamefont
  {Cortijo}}, \bibinfo {author} {\bibfnamefont {D.}~\bibnamefont {Kharzeev}},
  \bibinfo {author} {\bibfnamefont {K.}~\bibnamefont {Landsteiner}}, \ and\
  \bibinfo {author} {\bibfnamefont {M.~A.}\ \bibnamefont {Vozmediano}},\
  }\href@noop {} {\bibfield  {journal} {\bibinfo  {journal} {Phys. Rev. B}\
  }\textbf {\bibinfo {volume} {94}},\ \bibinfo {pages} {241405} (\bibinfo
  {year} {2016}{\natexlab{a}})}\BibitemShut {NoStop}%
\bibitem [{\citenamefont {Arjona}\ and\ \citenamefont
  {Vozmediano}(2018)}]{vozmediano2017-1}%
  \BibitemOpen
  \bibfield  {author} {\bibinfo {author} {\bibfnamefont {V.}~\bibnamefont
  {Arjona}}\ and\ \bibinfo {author} {\bibfnamefont {M.~A.}\ \bibnamefont
  {Vozmediano}},\ }\href@noop {} {\bibfield  {journal} {\bibinfo  {journal}
  {Phys. Rev. B}\ }\textbf {\bibinfo {volume} {97}},\ \bibinfo {pages} {201404}
  (\bibinfo {year} {2018})}\BibitemShut {NoStop}%
\bibitem [{\citenamefont {Pikulin}, \citenamefont {Chen},\ and\ \citenamefont
  {Franz}(2016)}]{Pikulin-2016}%
  \BibitemOpen
  \bibfield  {author} {\bibinfo {author} {\bibfnamefont {D.}~\bibnamefont
  {Pikulin}}, \bibinfo {author} {\bibfnamefont {A.}~\bibnamefont {Chen}}, \
  and\ \bibinfo {author} {\bibfnamefont {M.}~\bibnamefont {Franz}},\
  }\href@noop {} {\bibfield  {journal} {\bibinfo  {journal} {Phys. Rev. X}\
  }\textbf {\bibinfo {volume} {6}},\ \bibinfo {pages} {041021} (\bibinfo {year}
  {2016})}\BibitemShut {NoStop}%
\bibitem [{\citenamefont {Grushin}\ \emph {et~al.}(2016)\citenamefont
  {Grushin}, \citenamefont {Venderbos}, \citenamefont {Vishwanath},\ and\
  \citenamefont {Ilan}}]{Grushin-2016}%
  \BibitemOpen
  \bibfield  {author} {\bibinfo {author} {\bibfnamefont {A.~G.}\ \bibnamefont
  {Grushin}}, \bibinfo {author} {\bibfnamefont {J.~W.}\ \bibnamefont
  {Venderbos}}, \bibinfo {author} {\bibfnamefont {A.}~\bibnamefont
  {Vishwanath}}, \ and\ \bibinfo {author} {\bibfnamefont {R.}~\bibnamefont
  {Ilan}},\ }\href@noop {} {\bibfield  {journal} {\bibinfo  {journal} {Phys.
  Rev. X}\ }\textbf {\bibinfo {volume} {6}},\ \bibinfo {pages} {041046}
  (\bibinfo {year} {2016})}\BibitemShut {NoStop}%
\bibitem [{\citenamefont {Liu}, \citenamefont {Pikulin},\ and\ \citenamefont
  {Franz}(2017)}]{Liu-2017}%
  \BibitemOpen
  \bibfield  {author} {\bibinfo {author} {\bibfnamefont {T.}~\bibnamefont
  {Liu}}, \bibinfo {author} {\bibfnamefont {D.}~\bibnamefont {Pikulin}}, \ and\
  \bibinfo {author} {\bibfnamefont {M.}~\bibnamefont {Franz}},\ }\href@noop {}
  {\bibfield  {journal} {\bibinfo  {journal} {Phys. Rev. B}\ }\textbf {\bibinfo
  {volume} {95}},\ \bibinfo {pages} {041201} (\bibinfo {year}
  {2017})}\BibitemShut {NoStop}%
\bibitem [{\citenamefont {Arjona}, \citenamefont {Castro},\ and\ \citenamefont
  {Vozmediano}(2017)}]{Arjona-2017}%
  \BibitemOpen
  \bibfield  {author} {\bibinfo {author} {\bibfnamefont {V.}~\bibnamefont
  {Arjona}}, \bibinfo {author} {\bibfnamefont {E.~V.}\ \bibnamefont {Castro}},
  \ and\ \bibinfo {author} {\bibfnamefont {M.~A.~H.}\ \bibnamefont
  {Vozmediano}},\ }\href@noop {} {\bibfield  {journal} {\bibinfo  {journal}
  {Phys. Rev. B}\ }\textbf {\bibinfo {volume} {96}},\ \bibinfo {pages} {081110}
  (\bibinfo {year} {2017})}\BibitemShut {NoStop}%
\bibitem [{\citenamefont {Lundgren}, \citenamefont {Laurell},\ and\
  \citenamefont {Fiete}(2014)}]{Lundgren_014}%
  \BibitemOpen
  \bibfield  {author} {\bibinfo {author} {\bibfnamefont {R.}~\bibnamefont
  {Lundgren}}, \bibinfo {author} {\bibfnamefont {P.}~\bibnamefont {Laurell}}, \
  and\ \bibinfo {author} {\bibfnamefont {G.~A.}\ \bibnamefont {Fiete}},\
  }\href@noop {} {\bibfield  {journal} {\bibinfo  {journal} {Phys. Rev. B}\
  }\textbf {\bibinfo {volume} {90}},\ \bibinfo {pages} {165115} (\bibinfo
  {year} {2014})}\BibitemShut {NoStop}%
\bibitem [{\citenamefont {Jia}\ \emph {et~al.}(2016)\citenamefont {Jia},
  \citenamefont {Li}, \citenamefont {Shi}, \citenamefont {Liao}, \citenamefont
  {Yu},\ and\ \citenamefont {Wu}}]{Jia_016}%
  \BibitemOpen
  \bibfield  {author} {\bibinfo {author} {\bibfnamefont {Z.}~\bibnamefont
  {Jia}}, \bibinfo {author} {\bibfnamefont {C.}~\bibnamefont {Li}}, \bibinfo
  {author} {\bibfnamefont {J.}~\bibnamefont {Shi}}, \bibinfo {author}
  {\bibfnamefont {Z.}~\bibnamefont {Liao}}, \bibinfo {author} {\bibfnamefont
  {D.}~\bibnamefont {Yu}}, \ and\ \bibinfo {author} {\bibfnamefont
  {X.}~\bibnamefont {Wu}},\ }\href@noop {} {\bibfield  {journal} {\bibinfo
  {journal} {Nat. Commun.}\ }\textbf {\bibinfo {volume} {7}},\ \bibinfo {pages}
  {13013} (\bibinfo {year} {2016})}\BibitemShut {NoStop}%
\bibitem [{\citenamefont {Skinner}\ and\ \citenamefont
  {Fu}(2018)}]{Skinner_018}%
  \BibitemOpen
  \bibfield  {author} {\bibinfo {author} {\bibfnamefont {B.}~\bibnamefont
  {Skinner}}\ and\ \bibinfo {author} {\bibfnamefont {L.}~\bibnamefont {Fu}},\
  }\href@noop {} {\bibfield  {journal} {\bibinfo  {journal} {Sci. Adv.}\
  }\textbf {\bibinfo {volume} {4}},\ \bibinfo {pages} {eaat2621} (\bibinfo
  {year} {2018})}\BibitemShut {NoStop}%
\bibitem [{\citenamefont {Mu{\~n}oz}\ and\ \citenamefont
  {Soto-Garrido}(2017)}]{Munoz-2017}%
  \BibitemOpen
  \bibfield  {author} {\bibinfo {author} {\bibfnamefont {E.}~\bibnamefont
  {Mu{\~n}oz}}\ and\ \bibinfo {author} {\bibfnamefont {R.}~\bibnamefont
  {Soto-Garrido}},\ }\href@noop {} {\bibfield  {journal} {\bibinfo  {journal}
  {J. Phys.: Condens. Matter}\ }\textbf {\bibinfo {volume} {29}},\ \bibinfo
  {pages} {445302} (\bibinfo {year} {2017})}\BibitemShut {NoStop}%
\bibitem [{\citenamefont {Cortijo}\ \emph
  {et~al.}(2016{\natexlab{b}})\citenamefont {Cortijo}, \citenamefont
  {Ferreiros}, \citenamefont {Landsteiner},\ and\ \citenamefont
  {Vozmediano}}]{Cortijo-2016b}%
  \BibitemOpen
  \bibfield  {author} {\bibinfo {author} {\bibfnamefont {A.}~\bibnamefont
  {Cortijo}}, \bibinfo {author} {\bibfnamefont {Y.}~\bibnamefont {Ferreiros}},
  \bibinfo {author} {\bibfnamefont {K.}~\bibnamefont {Landsteiner}}, \ and\
  \bibinfo {author} {\bibfnamefont {M.~A.}\ \bibnamefont {Vozmediano}},\
  }\href@noop {} {\bibfield  {journal} {\bibinfo  {journal} {2{D} Materials}\
  }\textbf {\bibinfo {volume} {3}},\ \bibinfo {pages} {011002} (\bibinfo {year}
  {2016}{\natexlab{b}})}\BibitemShut {NoStop}%
\bibitem [{\citenamefont {Sakurai}\ and\ \citenamefont
  {Napolitano}(2014)}]{sakurai}%
  \BibitemOpen
  \bibfield  {author} {\bibinfo {author} {\bibfnamefont {J.~J.}\ \bibnamefont
  {Sakurai}}\ and\ \bibinfo {author} {\bibfnamefont {J.~J.}\ \bibnamefont
  {Napolitano}},\ }\href@noop {} {\emph {\bibinfo {title} {Modern Quantum
  Mechanics}}}\ (\bibinfo  {publisher} {Pearson Higher Ed},\ \bibinfo {year}
  {2014})\BibitemShut {NoStop}%
\bibitem [{\citenamefont {Neupane}\ \emph {et~al.}(2014)\citenamefont
  {Neupane}, \citenamefont {Xu}, \citenamefont {Sankar}, \citenamefont
  {Alidoust}, \citenamefont {Bian}, \citenamefont {Liu}, \citenamefont
  {Belopolski}, \citenamefont {Chang}, \citenamefont {Jeng}, \citenamefont
  {Lin} \emph {et~al.}}]{neupane-2014}%
  \BibitemOpen
  \bibfield  {author} {\bibinfo {author} {\bibfnamefont {M.}~\bibnamefont
  {Neupane}}, \bibinfo {author} {\bibfnamefont {S.-Y.}\ \bibnamefont {Xu}},
  \bibinfo {author} {\bibfnamefont {R.}~\bibnamefont {Sankar}}, \bibinfo
  {author} {\bibfnamefont {N.}~\bibnamefont {Alidoust}}, \bibinfo {author}
  {\bibfnamefont {G.}~\bibnamefont {Bian}}, \bibinfo {author} {\bibfnamefont
  {C.}~\bibnamefont {Liu}}, \bibinfo {author} {\bibfnamefont {I.}~\bibnamefont
  {Belopolski}}, \bibinfo {author} {\bibfnamefont {T.-R.}\ \bibnamefont
  {Chang}}, \bibinfo {author} {\bibfnamefont {H.-T.}\ \bibnamefont {Jeng}},
  \bibinfo {author} {\bibfnamefont {H.}~\bibnamefont {Lin}},  \emph {et~al.},\
  }\href@noop {} {\bibfield  {journal} {\bibinfo  {journal} {Nat. Commun.}\
  }\textbf {\bibinfo {volume} {5}},\ \bibinfo {pages} {3786} (\bibinfo {year}
  {2014})}\BibitemShut {NoStop}%
\bibitem [{\citenamefont {Datta}(1995)}]{Datta_1}%
  \BibitemOpen
  \bibfield  {author} {\bibinfo {author} {\bibfnamefont {S.}~\bibnamefont
  {Datta}},\ }\href@noop {} {\emph {\bibinfo {title} {Electronic Transport in
  Mesoscopic Systems}}}\ (\bibinfo  {publisher} {Cambridge Univ. Press},\
  \bibinfo {address} {Cambridge, U.K.},\ \bibinfo {year} {1995})\BibitemShut
  {NoStop}%
\bibitem [{\citenamefont {Datta}(2005)}]{Datta_2}%
  \BibitemOpen
  \bibfield  {author} {\bibinfo {author} {\bibfnamefont {S.}~\bibnamefont
  {Datta}},\ }\href@noop {} {\emph {\bibinfo {title} {Quantum Transport: Atom
  to Transistor}}}\ (\bibinfo  {publisher} {Cambridge Univ. Press},\ \bibinfo
  {address} {Cambridge, U.K.},\ \bibinfo {year} {2005})\BibitemShut {NoStop}%
\bibitem [{\citenamefont {Nazarov}\ and\ \citenamefont
  {Blanter}(2009)}]{Nazarov}%
  \BibitemOpen
  \bibfield  {author} {\bibinfo {author} {\bibfnamefont {Y.~V.}\ \bibnamefont
  {Nazarov}}\ and\ \bibinfo {author} {\bibfnamefont {Y.~M.}\ \bibnamefont
  {Blanter}},\ }\href@noop {} {\emph {\bibinfo {title} {Quantum Transport:
  Introduction to Nanoscience}}}\ (\bibinfo  {publisher} {Cambridge Univ.
  Press},\ \bibinfo {address} {Cambridge, U.K.},\ \bibinfo {year}
  {2009})\BibitemShut {NoStop}%
\bibitem [{\citenamefont {Ziman}(1960)}]{Ziman}%
  \BibitemOpen
  \bibfield  {author} {\bibinfo {author} {\bibfnamefont {J.~M.}\ \bibnamefont
  {Ziman}},\ }\href@noop {} {\emph {\bibinfo {title} {Electrons and Phonons:
  The Theory of Transport Phenomena in Solids}}}\ (\bibinfo  {publisher}
  {Oxford Univ. Press},\ \bibinfo {address} {New York},\ \bibinfo {year}
  {1960})\BibitemShut {NoStop}%
\bibitem [{\citenamefont {Madelung}(1978)}]{Madelung}%
  \BibitemOpen
  \bibfield  {author} {\bibinfo {author} {\bibfnamefont {O.}~\bibnamefont
  {Madelung}},\ }\href@noop {} {\emph {\bibinfo {title} {Introduction to
  Solid-State Theory}}}\ (\bibinfo  {publisher} {Springer},\ \bibinfo {address}
  {New York},\ \bibinfo {year} {1978})\BibitemShut {NoStop}%
\bibitem [{\citenamefont {Mu$\tilde{\text{n}}$oz}, \citenamefont {Lu},\ and\
  \citenamefont {Yakobson}(2010)}]{Munoz_010}%
  \BibitemOpen
  \bibfield  {author} {\bibinfo {author} {\bibfnamefont {E.}~\bibnamefont
  {Mu$\tilde{\text{n}}$oz}}, \bibinfo {author} {\bibfnamefont {J.}~\bibnamefont
  {Lu}}, \ and\ \bibinfo {author} {\bibfnamefont {B.~I.}\ \bibnamefont
  {Yakobson}},\ }\href@noop {} {\bibfield  {journal} {\bibinfo  {journal} {Nano
  Lett.}\ }\textbf {\bibinfo {volume} {10}},\ \bibinfo {pages} {1652} (\bibinfo
  {year} {2010})}\BibitemShut {NoStop}%
\bibitem [{\citenamefont {Mu$\tilde{\text{n}}$oz}(2016)}]{Munoz_CRC_016}%
  \BibitemOpen
  \bibfield  {author} {\bibinfo {author} {\bibfnamefont {E.}~\bibnamefont
  {Mu$\tilde{\text{n}}$oz}},\ }in\ \href@noop {} {\emph {\bibinfo {booktitle}
  {CRC Graphene Science Handbook: Optical and Electrical Properties}}},\
  Vol.~\bibinfo {volume} {3}\ (\bibinfo  {publisher} {CRC Press, Taylor {\&}
  Francis Group},\ \bibinfo {year} {2016})\ Chap.~\bibinfo {chapter} {17}, pp.\
  \bibinfo {pages} {253--271}\BibitemShut {NoStop}%
\bibitem [{\citenamefont {Mu$\tilde{\text{n}}$oz}(2012)}]{Munoz_012}%
  \BibitemOpen
  \bibfield  {author} {\bibinfo {author} {\bibfnamefont {E.}~\bibnamefont
  {Mu$\tilde{\text{n}}$oz}},\ }\href@noop {} {\bibfield  {journal} {\bibinfo
  {journal} {J. Phys.: Cond. Matt.}\ }\textbf {\bibinfo {volume} {24}},\
  \bibinfo {pages} {195302} (\bibinfo {year} {2012})}\BibitemShut {NoStop}%
\bibitem [{\citenamefont {Kirchner}, \citenamefont {Zamani},\ and\
  \citenamefont {Mu$\tilde{\text{n}}$oz}(2012)}]{Kirchner_12}%
  \BibitemOpen
  \bibfield  {author} {\bibinfo {author} {\bibfnamefont {S.}~\bibnamefont
  {Kirchner}}, \bibinfo {author} {\bibfnamefont {F.}~\bibnamefont {Zamani}}, \
  and\ \bibinfo {author} {\bibfnamefont {E.}~\bibnamefont
  {Mu$\tilde{\text{n}}$oz}},\ }\enquote {\bibinfo {title} {New {M}aterials for
  {T}hermoelectric {A}pplications: {T}heory and {E}xperiment},}\ \ (\bibinfo
  {publisher} {Springer},\ \bibinfo {year} {2012})\ Chap.~\bibinfo {chapter}
  {10}\BibitemShut {NoStop}%
\bibitem [{\citenamefont {Mu$\tilde{\text{n}}$oz}, \citenamefont {Kirchner},\
  and\ \citenamefont {Bolech}(2013)}]{Munoz_013}%
  \BibitemOpen
  \bibfield  {author} {\bibinfo {author} {\bibfnamefont {E.}~\bibnamefont
  {Mu$\tilde{\text{n}}$oz}}, \bibinfo {author} {\bibfnamefont {S.}~\bibnamefont
  {Kirchner}}, \ and\ \bibinfo {author} {\bibfnamefont {C.~J.}\ \bibnamefont
  {Bolech}},\ }\href@noop {} {\bibfield  {journal} {\bibinfo  {journal} {Phys.
  Rev. Lett.}\ }\textbf {\bibinfo {volume} {110}},\ \bibinfo {pages} {016601}
  (\bibinfo {year} {2013})}\BibitemShut {NoStop}%
\bibitem [{\citenamefont {Low}\ and\ \citenamefont {Guinea}(2010)}]{Low_010}%
  \BibitemOpen
  \bibfield  {author} {\bibinfo {author} {\bibfnamefont {T.}~\bibnamefont
  {Low}}\ and\ \bibinfo {author} {\bibfnamefont {F.}~\bibnamefont {Guinea}},\
  }\href@noop {} {\bibfield  {journal} {\bibinfo  {journal} {Nano Lett.}\
  }\textbf {\bibinfo {volume} {10}},\ \bibinfo {pages} {3551} (\bibinfo {year}
  {2010})}\BibitemShut {NoStop}%
\bibitem [{\citenamefont {Settnes}\ \emph {et~al.}(2016)\citenamefont
  {Settnes}, \citenamefont {Power}, \citenamefont {Brandbyge},\ and\
  \citenamefont {Jauho}}]{Settnes_016}%
  \BibitemOpen
  \bibfield  {author} {\bibinfo {author} {\bibfnamefont {M.}~\bibnamefont
  {Settnes}}, \bibinfo {author} {\bibfnamefont {S.~R.}\ \bibnamefont {Power}},
  \bibinfo {author} {\bibfnamefont {M.}~\bibnamefont {Brandbyge}}, \ and\
  \bibinfo {author} {\bibfnamefont {A.~P.}\ \bibnamefont {Jauho}},\ }\href@noop
  {} {\bibfield  {journal} {\bibinfo  {journal} {Phys. Rev. Lett.}\ }\textbf
  {\bibinfo {volume} {117}},\ \bibinfo {pages} {276801} (\bibinfo {year}
  {2016})}\BibitemShut {NoStop}%
\end{thebibliography}
\end{document}